\documentclass{bioinfo}

\usepackage[square,numbers,compress]{natbib}
\usepackage{amsthm}
\usepackage{amsmath}
\usepackage{todonotes}
\usepackage[normalem]{ulem}
\usepackage{url}
\usepackage{adjustbox}
\usepackage{lscape}
\usepackage{subcaption}
\newtheorem{definition}{Definition}
\newcommand{\mpara}[1]{\medskip\noindent{\bf #1}}

\usepackage{empheq}
\usepackage{tcolorbox}

\newtcbox{\mymath}[1][]{%
    nobeforeafter, math upper, tcbox raise base,
    enhanced, colframe=blue!30!black,
    colback=blue!30, boxrule=1pt,
    #1}

\copyrightyear{2021} \pubyear{2021}

\access{Advance Access Publication Date: Day Month Year}
\appnotes{Manuscript Category}

\begin{document}

\firstpage{1}

\subtitle{}

\title[A Review of Anonymization for Healthcare Data]{A Review of Anonymization for Healthcare Data}
\author[Olatunji \textit{et~al}.]{Iyiola E. Olatunji\,$^{\text{\sfb 1,}*}$, Jens Rauch\,$^{\text{\sfb 2}}$, Matthias Katzensteiner\,$^{\text{\sfb 3}}$, and Megha Khosla\,$^{\text{\sfb1}}$}
\address{$^{\text{\sf 1}}$L3S Research Center, Leibniz University, Hannover, Germany  \\
$^{\text{\sf 2}}$Health Informatics Research Group, University of Applied Sciences Osnabr\"{u}ck, Germany  \\
$^{\text{\sf 3}}$University of Applied Sciences and Arts Hannover, Germany.}

\corresp{$^\ast$To whom correspondence should be addressed.}

\history{Received on XXXXX; revised on XXXXX; accepted on XXXXX}

\editor{Associate Editor: XXXXXXX}

\abstract{
Mining health data can lead to faster medical decisions, improvement in the quality of treatment, disease prevention, reduced cost, and it drives innovative solutions within the healthcare sector. However, health data is highly sensitive and subject to regulations such as the General Data Protection Regulation (GDPR), which aims to ensure patient's privacy. 
Anonymization or removal of patient identifiable information, though the most conventional way, is the first important step to adhere to the regulations and incorporate privacy concerns. In this paper, we review the existing anonymization techniques and their applicability to various types (relational and graph-based) of health data. Besides, we provide an overview of possible attacks on anonymized data. We illustrate via a \emph{reconstruction attack} that anonymization though necessary, is not sufficient to address patient privacy and discuss methods for protecting against such attacks. Finally, we discuss tools that can be used to achieve anonymization.\\
\textbf{Availability:} Code is available: https://github.com/iyempissy/anonymization-reconstruction-attack.\\
\textbf{Contact:} \href{iyiola@l3s.de}{iyiola@l3s.de}
}

\maketitle

\section{Introduction}
\label{intro}

With the increasing adoption of healthcare information technology (HIT) by medical institutions, the generation and capture of healthcare-related data have been increasing rapidly in the past years. The application of artificial intelligence (AI) techniques already gives a glimpse of potential improvements ranging from lung cancer nodules detection in CT scans to disease prediction and treatment \cite{liu2018mtmr, pakbin2018prediction, tang2018predictive}. The challenge though is that these AI models are usually data hungry and require large amounts of data for training. Health care data, on the other hand, contains highly sensitive patient information and cannot be easily shared. 
The reluctance behind releasing data query/analysis tools build on health care data can be further justified by the fundamental law of information recovery \cite{CynthiaDwork10.1561/0400000042} which states that when a data source is queried multiple times and it returns overly accurate information for each query, the underlying data can be reconstructed partially or in full. Therefore, health data need to be protected against such leakage to ensure patient's privacy.

Privacy can be applied to health data at different levels. For instance, at the \emph{data collection phase}, randomization in the form of noise is usually added. Federated learning, homomorphic encryption, and secure multi-party computation can be applied at the \emph{data distribution phase}. 
In this work we focus on \emph{anonymization} which is used to achieve privacy at the \emph{data publication phase}. Consequently, regulations such as the GDPR \cite{officialgrpr2016} require data \emph{anonymization} or removal of personal or sensitive identification information before processing any knowledge extraction task or query. We provide a comprehensive review of anonymization in healthcare data focusing on the three main aspects: (i) anonymization models and techniques (ii) attacks and defenses proposed for anonymized data (iii) available tools for anonymizing data. 

In the first part, we introduce basic concepts and discuss existing anonymization techniques and their applicability to various types of health data. In particular, we differentiate between two different health data types: (i) relational and (ii) graph-based. Relational (same-site) patient data represents patient visits and medical diagnosis from a single hospital, and graph-based health data could include, for example, transmission network or epidemiological graph where the nodes are the patients and the edges are the interaction between the patients.

\subsection{Attacks on anonymization}
Although anonymized, the data might still be subject to several attacks such as \emph{background knowledge attacks}, \emph{linkage attacks}, \emph{attribute disclosure attacks}, and \emph{membership disclosure attacks}. A classic example of a linkage attack is the analysis of the 1990 U.S. Census performed by \citet{sweeney2000simple}. \citet{sweeney2000simple} found different combinations of pseudo or quasi-identifiers (QIDs) that would distinctively identify a person in the US and later used the same QID set to identify the then Massachusetts' governor (William Weld) as well as his medical record using a combination of information from an anonymous voter list and anonymized medical dataset obtained from the group insurance commission.

In the second part of the review, we focus on different attacks under different adversarial settings. For a practical illustration, we demonstrate by devising a reconstruction attack on the anonymized MIMIC-III dataset \cite{johnson2016mimic}. 

\subsection{Anonymization tools}
Finally, we review several existing tools which can be utilized by practitioners and researchers for anonymizing health data. 
These tools allow users to perform non-interactive anonymization on data. They can be integrated into popular database management systems (DBMS) such as MySQL, PostgreSQL, and Oracle, and can be used to generate synthetic dataset.

\subsection{Related works}
We differentiate our work from the following key surveys on anonymization \cite{hamza2013attacks, cite729gkoulalas2014publishing, cite7eze2015systematic, majeed2020anonymization}.
\citet{hamza2013attacks} only surveyed different attacks on privacy-preserving data publishing. \citet{cite729gkoulalas2014publishing} provided a comprehensive review on several algorithms for protecting health data. However, their survey does not cater for the tools or practical applicability of these methods. Besides, their review is only limited to relational data. \citet{cite7eze2015systematic} reviewed anonymization for health data but their survey is limited to data sharing. Moreover, their survey does not cater for different adversarial settings for which attacks can be successful on different methods.

Health data is dynamic (constantly changing) and has several attributes (high dimensional) which differentiate it from other types of data. However, the anonymization techniques provided in \cite{majeed2020anonymization} is generic to all types of data and not specific to health data. Therefore, most techniques reviewed cannot be directly applicable to health data.
Lastly, none of the aforementioned works showed the vulnerability of attack on a real-world dataset.

\subsection{Contributions}
We summarize our contributions as follows:
\begin{itemize}
    \item We present a comprehensive review of different anonymization models and data transformation techniques that have been applied to relational and graph-structured health data.
    \item We discuss different attacks on health data and demonstrate a practical reconstruction attack (code is provided at \url{https://github.com/iyempissy/anonymization-reconstruction-attack}) \\
    using the MIMIC-III dataset. We further review methods for protecting against such attacks.
    \item We highlight existing practical tools that can be used to preserve and analyze an individual's privacy from an adversary in healthcare settings. 
\end{itemize}
We believe that our work will assist researchers and practitioners in choosing appropriate anonymization techniques based on a multitude of aspects such as data type, desired privacy level, information loss, and possible adversarial behavior.
\subsection{Organization}
The rest of the paper is organized as follows. We present anonymization models (privacy models) and techniques for satisfying such models in Section \ref{sec:anonmodeltech}. We review methods for anonymizing different health data types in Section \ref{sec:diffdatatypes}. In Section \ref{sec:diffattacks}, we present attacks on anonymized data under different adversarial settings and demonstrate our reconstruction attack in Section \ref{sec:reconstructionattack}. Finally, we present the defense mechanism in Section \ref{sec:soaprivacypreservetech} and practical tools in Section \ref{sec:tools}.


\section{Anonymization models and techniques}
\label{sec:anonmodeltech}
\emph{Anonymization} refers to the complete removal of an individual's identifiers as well as the generalization of any other data that can be used to establish links to the individual. Though regulations like GDPR might require anonymization of data, a clear guideline does not exist. We remark that anonymization differs from de-identification which refers to the removal or replacement of  personal identifiers in the dataset such that the link between the individual and her data record can only be established by an authorized third party. 

Several privacy models and techniques for achieving anonymization have been proposed in the literature. Overall, the three main goals of anonymization are to preserve: data utility (measured by the amount of loss caused by the anonymization technique e.g information loss), privacy (measured by the conformity of the data to the privacy model constraints), and data truthfulness (each anonymized record corresponds to a single record in the original table) \cite{cite211fung2010introduction}. We start by giving an overview of proposed anonymization models in Section ~\ref{sec:anonmethods}, followed by techniques for operationalizing these models in Section \ref{sec:privacymodeltech}.

\subsection{Basic definitions and notations} 
The notion of privacy is often tied to the relational (tabular) data model \cite{cite211fung2010introduction}, in which (finite) datasets are organized in tables (relations) that consist of columns (attributes) and rows (records). We start by defining anonymization models for tabular data and later extend the notion to graph-based data in Section \ref{sec:graphanon}. First, we need the following notations and basic definitions.

Adhering to \cite{sweeney2000simple}, we denote tables as ${T}(A_1\dots,A_m)$, where the $\mathcal{A}=(A_k)_{1\leq k \leq m}$ are the attributes. For a record $t\in T$ and any subset of attributes $\{A'_j\}_{1 \leq j \leq r} \subset \mathcal{A}$, $r\leq m$ let $t[A'_1\dots,A'_r]$ be the sequence of values for a record with respect to the subset attributes $\{A'_j\}_{1 \leq j \leq r}$.

In the context of privacy, attributes can be categorized as direct identifiers (DIDs), quasi-identifiers (QIDs) and sensitive attributes (SAs) (cf. Table \ref{structureattrgraphattr}). DIDs uniquely identify an individual e.g social security numbers and driver's license. QIDs on their own cannot identify an individual but when combined, can re-identify the individual. SAs of an individual are to be kept private from potential adversaries, while non-sensitive attributes may be made public and are considered to be already known by adversaries. As QIDs form an important ingredient of various anonymization models, we formally define QID as follows:

\begin{definition}
Given a universe of individuals $\mathcal{U}$ and a table $\mathcal{T}(A_1,\dots,A_m)$ containing a dataset pertaining to a set of certain individuals, let $f: \mathcal{U} \rightarrow T$ and $g: T \rightarrow \mathcal{U'} \subset \mathcal{U}$. Then a \textbf{QID} is a set of non-sensitive attributes $Q_T = \{A'_j\}_{1\leq j \leq r} \subset \{A_1,\dots,A_m\}$ for which one individual can be re-identified when combined, i.e. $\exists p \in \mathcal{U}:g(f(p_i)[Q_T]) = p_i$.
\end{definition}

\subsection{Anonymization models}
\label{sec:anonmethods}
We here define three basic anonymization models proposed in the literature, namely, \textbf{$k$-anonymity}, \textbf{$\ell$-diversity}, and \textbf{$t$-closeness}. We also summarize various properties, limitations, and several variants of these models in Table \ref{tab:diffanonymization}. 

\subsubsection{$k$-anonymity} 
$k$-anonymity requires that at least $k$ individuals share the same attributes. Since QID contains fields that are likely to appear in other known datasets, $k$-anonymity ensures that each individual remains anonymous within their respective group (equivalence class).

\begin{definition}
Let $Q_T$ be the QID for table $\mathcal{T}(A_1,\dots,A_m)$, then $T$ satisfies \textbf{$k$-anonymity} if and only if there are at least $k$ identical records for each unique combination $t[Q_T]$ of records values with respect to the QID.
\end{definition}

For example, if $k$ = 10, then each equivalence class should have at least 10 similar records. This guarantees that the attacker cannot identify the identity of a single record. However, when $k$ is too high, utility depreciates. Moreover, the absence of sufficient heterogeneity in sensitive attributes limits the privacy offered by the $k$-anonymity model.

\subsubsection{$\ell$-diversity} 
$\ell$-diversity overcomes the limitations of $k$-anonymity by considering diversity among SAs. $\ell$-diversity ensures that there are at least $\ell$-distinct values of SA in each equivalence class \cite{cite3machanavajjhala2007diversity}.

Suppose, that for the QID value combination $q$, there exists an equivalence class $q^*$. The set of records $t$ in table $T$, whose values of $q$ all belong to $q^*$ is called a $q^*$-block.

\begin{definition}
\label{def:ldiversity}
A $q^*$-block is $\ell$-diverse if it contains at least $\ell$ different values for the SA $S$ and the $\ell$ most-frequent values have roughly the same frequency. A table is \textbf{$\ell$-diverse} if every $q^*$-block is $\ell$-diverse.
\end{definition}

However, $\ell$-diversity cannot prevent attribute disclosure attacks.

\mpara{\textit{Example.}} Assuming disease is the SA and that the table is 3-diverse (meaning 3 distinct SA values in each equivalence class). For an equivalence class where the attribute values are gastric ulcer, gastritis, and stomach cancer, an adversary can infer that an individual has stomach-related problems because all three diseases in the equivalence class are stomach-related.

\subsubsection{$t$-closeness} 
$t$-closeness ensures that the distance between the distribution of sensitive values in each equivalence class and the original class is no more than a threshold $t$ \cite{cite3li2007t}. Hence, a smaller value of $t$ represents stronger privacy.

\begin{definition}
\label{def:tcloseness}
Let $P_s$ be the distribution of SA $s$ in table $T$ and $Q_s$ be the distribution of this attribute in a $q^*$-block. The $q^*$-block is said to have \textbf{$t$-closeness} if the distance $d$ between $P_s$ and $Q_s$ is at most $t$ for any SA $s$. The table $T$ has $t$-closeness if this holds for all its $q^*$-blocks.
\end{definition}

$t$-closeness overcomes the limitation of $\ell$-diversity in presence of skewed attribute distributions in which one sensitive value dominates. Ensuring $t$-closeness for the above example would imply that for a table that satisfies 3-diversity, each equivalence class will not contain all stomach-related problems but other types of disease such as pneumonia. Common distance function used to measure closeness includes Kullback-Leibler (KL) distance and Earth Mover's Distance (EMD).

\subsection{Techniques for satisfying privacy models}
\label{sec:privacymodeltech}
Several techniques that have been used to satisfy various privacy models ($k$-anonymity, $\ell$-diversity, $t$-closeness) include slicing, generalization, suppression and relocation, perturbation, bucketization, and microaggregation. 

\mpara{Slicing.} Slicing involves partitioning the data into groups both horizontally and vertically. The slicing technique first performs vertical slicing by partitioning attributes into columns where each column contains a subset of attributes. Then performs horizontal slicing by partitioning tuples into buckets where each bucket contains a subset of tuples \cite{cite2li2010slicing}. The goal of slicing is to ensure that attributes that are highly correlated are grouped together. 

\begin{figure}[!tpb]
\centerline{\includegraphics[width=\linewidth]{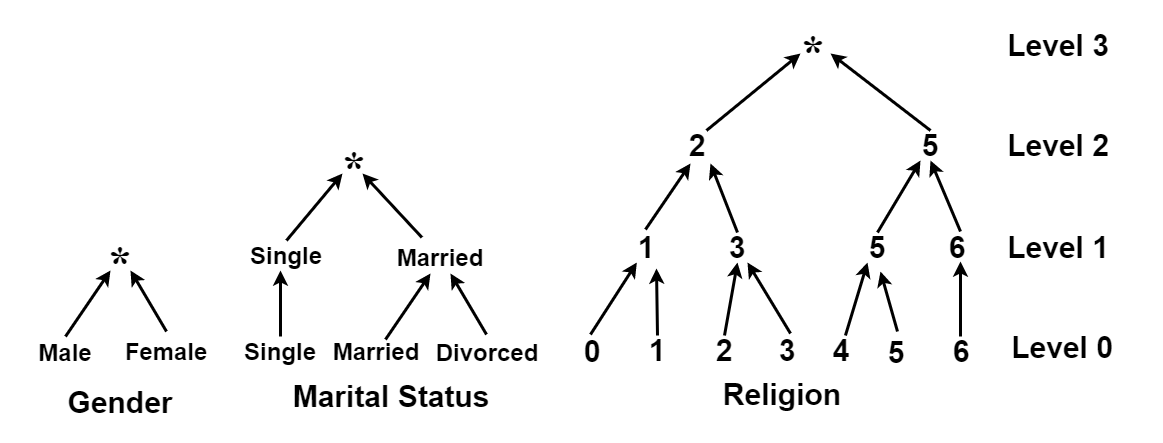}}
\caption{Example of generalization hierarchies of gender, marital status, and religion. Religion is label-encoded indicating Greek orthodox, protestant, others, Jewish, Methodist, catholic, and christian scientist. Level refers to the height of the hierarchy tree and \textbf{*} indicates suppression.}\label{generalizationhierarchy}
\end{figure}

\mpara{Generalization.} Generalization replaces QIDs (attributes that potentially identify individual i.\,e., age, zip code, gender, date of birth, etc.) values with other less specific values which are consistent with the original data \cite{cite2lee2017utility}. An example of generalization hierarchy for three attributes (gender, marital status, and religion) is shown in Figure \ref{generalizationhierarchy}.

Since generalization may lose considerable information, various methods have been proposed to reduce the information loss. 
Most generalization algorithms use global recoding or full domain generalization of attribute values. This implies that the same transformation is applied to each QID value. However, data utility can be higher when local recoding is applied to each equivalence class \cite{cite433xu2006utility}. Anonymization achieved with generalization-only approaches inevitably distorts the records. Therefore, over-generalization negatively influences the analysis of the anonymized dataset.

\mpara{Suppression and relocation.} 
In order to reduce over-generalization, suppression and relocation techniques are used \cite{cite28sweeney2002achieving, cite29nergiz2014hybrid}. Suppression involves the removal of outliers, and relocation involves the changing of the QIDs of the outliers. Outliers are the main cause of over-generalization because the outliers are distant from other records and since they are few, it is not sufficient to group them in the same equivalent classes. In these methods, the values of outliers are removed or changed.

Suppression can be performed at the record level or cell level. Record level suppression leads to excessive deletion of records for equivalence classes that do not meet privacy constraints while cell level suppression only removes QID values that make a tuple violate privacy model constraint. Thereby leading to reduced loss of information. Suppression comes in handy when used together with machine learning models because the suppressed values can be handled as missing values. However, in large datasets, full column suppression can potentially hurt utility. 

\mpara{Perturbation.} In lieu of suppression, perturbation techniques can be used to augment generalization. Perturbation involves replacing sensitive values and QIDs with fake masks of the original data \cite{cite7eze2015systematic}. Using perturbation to achieve anonymization leads to better utility because the degree of generalization is limited via the insertion of counterfeit records to equivalent classes \cite{cite2lee2017utility}. 

\mpara{Bucketization.} Bucketization \cite{cite2aggarwal2005k, cite2kifer2006injecting} separates the sensitive attributes from the QIDs by randomly permuting or swapping the sensitive attribute values in each bucket. This achieves better utility than generalization. However, because bucketization publishes the QID values in their original forms, an adversary can still identify an individual using the QIDs and therefore, does not prevent membership disclosure. For better privacy, generalization and bucketization can be applied on the same dataset to jointly achieve $k$-anonymity and $\ell$-diversity respectively.

\mpara{Microaggregation:} Ensuring that the same dataset satisfies $k$-anonymity, $\ell$-diversity and $t$-closeness leads to considerable loss of information due to generalization, perturbation and suppression or may not be even achievable in practice. Thus, \citet{cite727gal2014data} proposed a microaggregation algorithm that caters for numerical QIDs for creating $k$-anonymous equivalent classes by replacing sensitive attributes with masked values. \cite{cite27kieseberg2016tamper}.
Microaggregation involves substituting the values of groups of $k$ nearest records by their centroid. The optimal solution to this problem is known to be NP-hard \cite{cite5oganian2001complexity, cite7domingo2005ordinal}.

\begin{table}[!h]
\caption{Attributes of different health data types.}
\label{structureattrgraphattr}
\begin{tabular}{p{2.5cm}p{5cm}}
Attributes & Description \\ \toprule
\multicolumn{2}{l}{\textbf{Relational data attributes}} \\ \midrule
Direct identifier (DID) & Identifiers that uniquely identifies an individual e.g social security numbers and driver's license. \\
Quasi-identifier (QID) & QIDs on their own cannot identify an individual but when combined, can re-identify the individual e.g gender, postcode \\
Sensitive attributes (SA) & SAs are not usually public data but are sensitive if associated with an individual e.g drug codes, diagnosis, and disease conditions \\ \midrule
\multicolumn{2}{l}{\textbf{Graph-based data attributes}} \\ \midrule
Node & Represents individuals. Even when the PII are removed from the nodes, the presence or absence of a target individual in the graph can be considered as the privacy of the individual. \\
Node properties & Node properties such as the degree (number of connected neighbors) or attributes associated with a node e.g sex can be considered as privacy of the individual \\
Node labels & The labels of nodes in a graph can be non-sensitive labels or sensitive labels. The sensitive labels are similar to the sensitive attributes in relational data \\
Links & The links or edges in graphs show that there exists a relationship between corresponding nodes. Therefore, needs to be kept private
\end{tabular}
\end{table}

\section{ Anonymization in healthcare: data types and techniques}
\label{sec:diffdatatypes}
\subsection{Relational health data} 
The most conventional way of representing patients' health data is relational(tabular) where each row (a record) in a database table represents a measurement, event or other patient-related information. This data model is known as relational and typically represents single-site patient data, which includes patient visits and medical diagnosis from a single hospital or doctor's practice.

For example, when patient $A$ visits a doctor in the same hospital, the diagnosis and details of $A$'s visits form a record in the database. This forms the basis of the electronic medical records (EMR). For this type of health data, anonymization is achieved by ensuring that the records in each equivalence class are indistinguishable or similar.

\subsubsection{Anonymization of relational health data}
\label{sec:structuredanon}
\mpara{Anonymization based on generalization.}
Following the definitions in Section \ref{sec:anonmethods}, the combination of QIDs such as age, race, gender, and zip code can be used by an adversary to re-identify an individual. Hence, QIDs are usually generalized to satisfy the constraints of the privacy models.

To reduce information loss due to generalization, the $k$-anonymity problem can be viewed as a clustering or partitioning problem where the goal is to find sets of clusters (i.e., equivalence classes), each of which contains at least $k$ records \cite{cite6byun2007efficient, cite63agrawal2003information, cite68dong2005reference, cite733mortazavi2014fast, cite735nergiz2008multirelational}. 
The records in each partition or cluster are then generalized so that they all share the same QID value and have at least $k$-records. However, when there are many QIDs, the clustering technique still suffers from high information loss \cite{cite723aggarwal2008privacy, cite734narayanan2008robust} and fails to protect against $\ell$-diversity.

While most of the works ignored the differences of QIDs' potential to reveal an individual's identity,
\citet{cite4majeed2017vulnerability} focused on adaptive attribute generalization based on different information contents of QIDs. They proposed a solution to $\ell$-diverse and $t$-closeness in an imbalance dataset by creating a flexible generalization hierarchy based on the vulnerability of the QIDs to reveal the identity and the diversity of SAs in each equivalence class. Their proposed method enhances utility by controlling the over-generalization of QIDs that are less vulnerable to revealing identity in diverse classes.

For the same dataset to be $k$-anonymous, $\ell$-diverse, and $t$-close, \citet{cite727gal2014data} proposed a microaggregation algorithm that utilizes numerical QIDs to create $k$-anonymous clusters and replaces sensitive attributes with masked values. Other methods used to jointly achieve all $k$-anonymity, $\ell$-diversity, and $t$-closeness on medical data include SHARE \cite{cite729gkoulalas2014publishing} and correlation-aware anonymization of high-dimensional data (CAHD) \cite{cite728ghinita2008anonymization}. 

\mpara{Anonymization for incremental medical data.} Medical records are incremental in nature, therefore, methods for incremental anonymization or data streams have been proposed where the goal is to anonymize medical data as they increase \cite{cite724byun2006secure}, \cite{cite726fung2011service}, \cite{cite79kim2014framework, cite737pei2007maintaining, cite739wu2009privacy}, \cite{cite7mohammadian2014fast}, \cite{cite7kim2018privacy}, \cite{cite7otgonbayar2018k}, \cite{cite7otgonbayar2019x}. Such methods are based on accumulation, aggregation, and clustering. For these methods, each new piece of data is a tuple consisting of QIDs and sensitive attributes. When new data arrive, they pass through an aggregation engine that assigns the data to a cluster based on the information loss. These clusters are then passed to the privacy models. If a tuple does not satisfy the privacy model e.g $\ell$-diversity, perturbation is applied to the tuple by randomly adding some generated attributes to the original data.

\mpara{Utility-aware anonymization.} Utility-aware anonymization was considered in \cite{cite432lefevre2006mondrian, cite433xu2006utility, POULIS201776}. These methods use attribute-level anonymization that retains the original values of QIDs based on the assigned utility value and applies local recoding to anonymize the data. Moreover, \citet{POULIS201776} proposed a ($k$, $k^m$)-anonymity technique that has bounded utility constraints and still ensures the privacy of both demographic information and disease diagnosis codes.

As knowledge of the interrelationship between QIDs and SAs affects privacy protection \cite{cite435li2008injector, cite436du2008privacy, cite437wong2007minimality, cite438tao2008anti}, a number of works focused on determining the possible level of privacy without compromising utility. Besides using popular metrics such as information loss, precision, and discernability to measure utility, several works used the accuracy of a trained classification model as the utility metric. Such approaches lean towards employing machine learning (ML) models or decision trees to measure the accuracy-anonymization/privacy trade-off  \cite{cite427zaman2016novel, cite441lefevre2006workload, cite442fung2007anonymizing, cite443li2011information, cite444kisilevich2009efficient}.

\mpara{Anonymizing multi-dimensional health data.} Since health data may be multi-dimensional with several attributes, the knowledge of an attacker knowing those attributes can be relaxed and bounded to $L$ according to the $LKC$-privacy model proposed by \citet{cite732mohammed2010centralized}. This implies that not all attributes need to be anonymized but a combination of $L$ attributes in order to satisfy the privacy model (e.g. $k$-anonymity). However, combinatorics suffers from the curse of dimensionality \cite{cite723aggarwal2008privacy} and the anonymized dataset can still be vulnerable to inference attacks when viewed collectively \cite{cite724byun2006secure}. Therefore, data publishers should anonymize and publish only a sample of the original dataset \cite{cite725el2008protecting}.

\mpara{Anonymizing against background attack.} \citet{cite3majeed2019attribute} proposed an anonymization scheme that prevents individuals from identity disclosure even when faced with adversaries having strong background knowledge. Their proposed method is based on transforming data into fixed intervals and then replacing the original values with averages.

\subsection{Graph-based health data} 
With the adoption of electronic health record (EHR) which allows cross-sectional as well as the longitudinal view of patient diagnosis across different hospitals (multi-site), several works have been geared towards representing EHR data in combination with other heterogeneous sources such as clinical notes, diagnosis (ICD codes), therapies and prescriptions (procedure and ATC codes) as graphs  \cite{bearman2004chains, ji2016graph, shi2017semantic, gyrard2018personalized, cong2018constructing, yu2017knowledge}. \citet{liu2016graph} used graph analysis technique to detect fraud, waste, and abuse in healthcare by representing health dataset as a heterogeneous network consisting of several patients, doctors, and pharmacies. Similarly, \citet{branting2016graph} used graphs derived from an open-source health dataset for estimating healthcare fraud. Recently, EHRs have been represented as graph data to extract knowledge graphs \cite{shi2017semantic, gyrard2018personalized, cong2018constructing, yu2017knowledge}.

These graph-based health data provide rich knowledge about the interaction and the underlying properties of the data. For example, in a disease transmission network or epidemiological graph, the nodes are the patients whereas the edges represent the interaction between the patients. Therefore, both nodes and edges need to be protected as they contain sensitive information. Anonymization approach for graph-based health data can be achieved either by removing edges and node labels (na\"{i}ve anonymization) or by adding new edges or nodes to modify the structure of the graph (structural anonymization).

\subsubsection{Anonymization of graph-based health data}
\label{sec:graphanon}
Most real-world data including health data can be represented as graphs where nodes denote entities or individual and edges represent the interaction or relationship between the entities 
To anonymize graphs, one can re-represent the data as a single table and apply anonymization as though it was relational data. However, graphs do not have the same semantics as relational data where records are independent and such representation will fail to provide the necessary anonymization. The graph-variant of the $k$-anonymity technique used in relational health data is $k$-degree anonymity. 

\begin{definition}
\label{def:kdegreeanonymity}
A graph is $k$-degree anonymous if for every node $v$, there exist at least $k$-1 other nodes in the graph with the same degree as $v$.
\end{definition}

\citet{cite518zhou2008preserving} showed that even when an individual's privacy is preserved with conventional anonymization techniques, the individual in a graph can be re-identified due to neighborhood attacks. This is possible when the adversary has some knowledge about the neighbors of a target victim and the relationship among the neighbors. To ensure privacy, each node must have $k$ other nodes with the same $1$-hop neighborhood.

\mpara{Anonymization based on graph manipulation.} Using graph modification technique such as edge addition or deletion, \citet{cite58liu2008towards} achieved $k$-degree anonymity by constructing a $k$-anonymous degree sequence where each node has the same degree as $k$-1 other nodes. The anonymous degree sequence is then used in constructing the anonymized graph. \citet{mortazavi2020gram} proposed ($k$, $\ell$)-anonymity method that preserves the privacy of the individual or node in the graph such that even when an attacker knows at most $\ell$ neighbors of a node, an attacker cannot identify that node in a group of less than $k$ nodes. This is achieved by first adding edges to the graph to satisfy the privacy model, then redundant edges are removed to minimize changes between the anonymized and the original graph. However, the changes may be visible to an attacker. Similarly, \cite{cite5zou2009k}, \cite{cite5cheng2010k}, \cite{cite5yuan2010personalized} and \cite{cite5andreou2017identity} exploited graph structure and attribute information to protect against such structure-based and node attribute disclosure attacks (see Section \ref{sec:graphattacks}) using $k$-degree anonymity, $\ell$-diversity, and $t$-closeness.

\mpara{Clustering-based anonymization.} Similar to the relational data in Section \ref{sec:structuredanon}, clustering-based anonymization techniques for graph data have been proposed \cite{bhagat2009class, hay2008resisting, thompson2009union}. These approaches first partition the data into clusters and construct graphs based on the data in each cluster. Then graph modification techniques such as removing or adding edges are performed on the constructed graph to satisfy the privacy model constraints. \citet{cite517zheleva2007preserving} showed that sensitive relationships can be inferred from anonymized graphs. As a defense method, they removed all edges from the anonymized graph and collapsed the nodes into a single node for each cluster. Then they randomly selected edges from the removed edges and assigns them as the edge of the cluster.
In order to anonymize graphs while maintaining structural similarity, \citet{foffano2019you} applied Szemer\'{e}di regularity lemma in partitioning each node into clusters and ensuring that the intra-partition edges behave almost randomly. By this, the inter-partition edges are left unaltered, making it semantically similar to the original graph.

Although we allude that several methods for anonymizing social network data might be applicable for graph-based health data, such claims have not been empirically verified. Interested readers are referred to \cite{majeed2020anonymization} for details about anonymizing social network data.

\begin{table*}
\begin{center}
\caption{Comparison of different privacy models, guarantees and their variants. The first 3 rows are  for relational data while the last 2 rows are for graph-based data.}
\label{tab:diffanonymization}
\begin{tabular}{p{1.5cm}p{2cm}p{4cm}p{5cm}p{4cm}}
\toprule
Privacy Model & Protected Attribute & Guarantee & Limitation & Variants \\ \toprule
$k$-anonymity & QID & Protection against LI/identity disclosure attack and AD attack. & Privacy is not guaranteed when an adversary has strong background knowledge (BK attack) or when the SA values in an equivalence class are not diverse (homogeneity attack). & ($k$, $e$)-anonymization \cite{zhang2007aggregate} 

$p$-sensitive $k$-anonymity \cite{campan2010p}

$p+$-sensitive $k$-anonymity and ($p$, $\alpha$)-sensitive $k$-anonymity \cite{sun2011extended}

$\theta$-sensitive $k$-anonymity \cite{khan2020theta}

($\alpha$, $k$)-anonymity \cite{wong2006alpha}

complete ($\alpha$,$k$)-anonymity \cite{jian2008complete}

SDSV-$p$-sensitive $k$-anonymity \cite{budiardjo2019approach}

($P$, $U$)-sensitive $k$-anonymity \cite{agarwal2018enhanced}

overlapped slicing method \cite{budiardjo2019privacy}

MSA($\alpha$,$\ell$)  algorithm \cite{zhang2017improved}

($k$,$k^m$)-anonymity \cite{POULIS201776}

DBTP-MDAV \cite{wu2019micro} \\

\midrule

$\ell$-diversity & SA & Protection against AD attack and homogeneity attack. Offers protection of SA. & Difficulty in creating a feasible $\ell$-diverse dataset when the data is highly imbalanced (e.g., if the SA distribution is uneven). Does not protect against MD attack. & independent $\ell$-diversity \cite{zhu2015privacy}

($\tau$, $\ell$)-diversity \cite{tian2011extending}

DBTP-$\ell$-MDAV \cite{wu2019micro}

($c$, $k$)-anonymization \cite{xiao2020privacy}

($\ell$, $d$)-semantic diversity \cite{oishi2020semantic}

SQ $\ell$-diversity \cite{sei2017anonymization} \\

\midrule
$t$-closeness & SA & Depending on the value of $t$, it may protect against MD attacks. Well suited for numeric attributes and offers better protection than $\ell$-diversity. & Difficult to create a feasible $t$-close dataset when data is highly imbalanced. A higher value of $t$ may lead to degradation in data utility and it is complex in nature. & ($n$,$t$)-closeness \cite{li2009closeness}

$M$-Shuffle based $t$-closeness \cite{qu2017privacy}

Multiple sensitive attribute-based $t$-closeness \cite{wang2018privacy}

SABRE $t$-closeness \cite{cao2011sabre}

Microaggregation-based $t$-closeness \cite{soria2015t}

SQ $t$-closeness \cite{sei2017anonymization} \\
\midrule
$k$-degree anonymity & Node and links & Protection against NAD attack, NB attack, and ST attack. & Vulnerable to NE attack. The edge additions and removal may adversely alter the structure of the original graph.  & $k$-anon graphs \cite{cite518zhou2008preserving}

$k$-degree anonymity \cite{cite58liu2008towards}

class-based anonymity \cite{bhagat2009class}

$(s,g)$ anonymity \cite{hay2008resisting}

cluster-based anonymity \cite{thompson2009union} \\
\midrule
$k$-degree-$\ell$-diversity & Node properties and labels & Protection against NE attack. Similarly, node labels cannot be inferred.  & The noise added to nodes to preserve privacy may lead to lower utility.  & $k$-automorphism \cite{cite5zou2009k}

$k$-isomorphism \cite{cite5cheng2010k}

$p$-anonymity \cite{cite5yuan2010personalized}

Gram algorithm \cite{mortazavi2020gram}

$s$L-anonymity \cite{foffano2019you} \\

\botrule
\end{tabular}
\end{center}
\end{table*}

\section{Attacks on anonymized health data under different adversarial settings}
\label{sec:diffattacks}
In this section, we discuss the different types of adversaries and plausible attacks on graph-based and relational health data. We also discuss the vulnerability of the attack under certain adversarial settings as summarized in Table \ref{advasaryclass}.

\subsection{Types of adversaries}
\label{sec:adversarytype}
When considering anonymization, understanding the knowledge of the adversary provides better protection against re-identification and also plays a crucial role when performing analysis \cite{cite730jandel2014decision}. An adversary can either be \textit{semi-honest} or \textit{malicious}. A semi-honest adversary (honest-but-curious) is an adversary that follows the predefined protocol but also interested in learning more from the received information than he is entitled to. A malicious adversary, however, deviates from the protocol and possibly colludes with other corrupted parties. We classify these adversaries into classical, statistical, and adaptive adversaries.

\begin{table}[h!]
\caption{Categorization of different types of adversaries and their attack vulnerability.}
\label{advasaryclass}
\begin{tabular}{p{1.2cm}p{3.1cm}p{1.7cm}p{1.3cm}}
\toprule
Adversary Type & Description & Categorization & Attacks \\ \toprule
Classical & Tries to discover SAs of an individual based on her knowledge of the QIDs. & Semi-honest & BK, NAD, LI  \\ 
\midrule

Statistical & Exploits the differences between the statistical distribution of the original and the anonymized dataset to uncover perturbations applied to the data. & Semi-honest & AD, MD, NE \\ 
\midrule

Adaptive & Has the capability to reverse-engineer the algorithm used for the anonymization based on her knowledge of the anonymization algorithm. Moreover, she has the capability of adapting her strategy as the attack progresses.& Malicious & AD, MD, NB, ST, NE \\
\botrule

\end{tabular}
\end{table}

\subsection{Attacks on anonymized relational data}
\label{sec:structuredattacks}
Following the adversarial settings in Section \ref{sec:adversarytype}, we categorize the attacks on relational health data into \emph{background knowledge attacks}, \emph{linkage attacks}, \emph{attribute disclosure attacks}, and \emph{membership disclosure attacks}.

\mpara{Background knowledge (BK) attack.} When an adversary knows some information or QIDs about the target individual, she can reconstruct the identifiable information of the individual. Such a reconstruction attack compromises the privacy of the target individual. We show an example of a reconstruction attack in Section \ref{sec:reconstructionattack} using the MIMIC-III dataset.

\mpara{Linkage (LI) attack.} The linkage attack is one of the classical attacks on relational data where an adversary can re-identify or link a record in an anonymized dataset by combining QIDs from different sources to an individual. This requires some form of background knowledge attack.

\mpara{Attribute disclosure (AD) attack.} In an AD attack, the  attacker aims to gain new information on SA. The attacker can also exploit the properties of the QIDs to estimate the SA. Usually, attribute disclosure is a by-product of identity disclosure. Nonetheless, it can be agnostic to linkage attack. For example, an adversary may be interested in inferring SA that is common to all target individuals.

\mpara{Membership disclosure (MD) attack.} MD attacks involve an adversary aiming to infer the presence or absence of an individual in a dataset. For example, knowing that an individual is present in a cancer dataset reveals that she has cancer even though the specific type of cancer might not be inferred. This may serve as leverage to launch an attribute disclosure attack.

\subsection{Attacks on anonymized graph data}
\label{sec:graphattacks}

On relational data, a combination of QIDs can be used in identifying the individual as in linkage attack. However, in graph-based data, several additional parameters can be used to uniquely identify an individual. These attributes include node, node properties or attributes, node labels, and links (see also Table \ref{structureattrgraphattr}). The combination of the attributes implies that several pieces of information can be considered as privacy of an individual which leads to several attacks that are different from relational health data.

\mpara{Node attribute disclosure (NAD) attack.} This attack is similar to the linkage attack in the relational data where QID can be used to re-identify an individual. A node may be linked to an individual based on the set of attributes (like age, gender, etc) assigned to the node. 

\mpara{Neighborhood (NB) attack.} Since nodes are connected in the graph, an adversary having background knowledge about some of the neighbors can uncover the identity of other connected nodes. For example, before the anonymized graph is released, an adversary can randomly create a subgraph and attach it to some target user. The attack is successful if the attacker can find the subgraph in the released anonymized graph. She can then discover other nodes by following the edges of the subgraph.

\mpara{Structure-based (ST) attack.} The ST attack is an umbrella type of attack consisting of several attacks where the adversary utilizes the unique structural characteristics of the graph. These attacks include degree attack (using the degree sequence of the graph to uniquely identify an individual), subgraph attack (the attacker knows the subgraph around the target node), hub-fingerprint attack (an adversary knowing the distance between a hub and target node can invade the privacy of the target node), and walk-based attack (searching over short walks in the graphs)

\mpara{Node existence (NE) attack.} Similar to the membership disclosure attack in the relationa data, the adversary aims to discover the presence or absence of a node in the graph \cite{olatunji2021membership}. This attack is usually preceded by the structure-based attack.

\begin{methods}
\section{Reconstruction attack}
\label{sec:reconstructionattack}
\subsection{Motivation}
\label{sec:motivation}
\citet{tang2018predictive} showed that demographic features such as insurance, marital status, gender, age, and race does not contribute to patient readmission prediction models. From a privacy perspective, this implies that demographics which are a form of QIDs that can be used to re-identify an individual when combined with other publicly available data can be safely removed without affecting the prediction result. Similarly, the remaining (non-private) data can be released without violating patient's privacy. 

In this paper, we show that this approach of releasing data still poses a threat to privacy. Assume that an adversary has some demographic information of some data records, she can train a classifier or ML model using the released data (with the demographics data removed) and the subset of demographic information that she has to predict the demographics of the entire dataset. That is, the demographic information of the rest of the patients, for whom she does not have can be inferred. This implies that demographics encodes some amount of information and in the presence of other attributes, demographics seems not to be useful (i.e. it is correlated with other attributes).

We allude that the assumption that the adversary can have access to demographic information of some of the patients is valid as, for example, some people post their diagnoses or symptoms online. For instance, \citet{guntuku2020tracking} showed that social media can be used for symptom discovery and diagnosis of diseases. They mapped tweets to states by geolocating all tweets using a combination of location coordinate information and user location descriptions. Similarly, demographics can be obtained from social media profiles which when mapped with the diagnosis can be used by an adversary to launch an attack. To summarize, we show that it does not suffice to only remove demographic information from data to ensure privacy.

\subsection{Dataset}
The Medical Information Mart for Intensive Care (MIMIC-III) \cite{johnson2016mimic} dataset consists of hospital admission data, lab measurements, procedure event recordings, prescriptions, hospital length of stay, diagnostic codes, and microbiological data of 53,467 unique patients. We extract demographics and data for our readmission prediction experiment using the same subset of the dataset as in \cite{pakbin2018prediction} while for the length of stay (LOS) prediction, we used the same subset of the dataset as in \cite{tang2018predictive}. We used age, ethnicity, admission type, marital status, insurance, religion, gender, and language as the demographic information. We grouped age into 5 equally spaced bins. $x\_19$ represents the features used in predicting the corresponding task. They include potassium score, arterial blood pressure, albumin score, blood urea nitrogen (BUN) score, creatinine score, sodium score, bicarbonate score, heart rate, systolic blood pressure, temperature, respiratory rate, spo2, glucose level, coagulation, physiology score (sapsii), pao2fio2 score, sirs, organ failure (sofa) and acute physiology score (apsiii).

\subsection{Methods}
\label{sec:methods}
We used the ARX tool \cite{prasser2020flexible} for implementing our privacy models. We first define multiple transformation models for each attributes based on generalization and suppression that fulfills the risk threshold of the privacy model as shown in Figure \ref{generalizationhierarchy}. This creates a hierarchy of possible solution space as shown in Figure \ref{solutionspacelattice}. The algorithm then searches the solution space for optimal solution (yellow colored node) according to the utility model. The utility model $\mathcal{U}_m$ measures the loss of information. It is defined as:
$$1-\frac{1}{m}\sum_{1\leq x \leq m} \mathcal{L}(x)$$
where $\mathcal{L}(x) \in [0,1]$ is the information loss for attribute $x$ defined as:
$$\mathcal{L}(x) = \frac{1}{n} \sum_{1 \leq y \leq n} loss(x,y)$$ 

where $m$ is the number of attributes and $n$ is the number of records in the dataset and $loss(x,y)$ is the information loss of a particular record $y$ of an attribute $x$. This depends on the attribute type.

For categorical and numeric attributes:
$$loss(x,y) = \frac{leafs(x, \mathcal{V}(x,y)) - 1}{leafs(x, root(x)) - 1}$$

For attributes with intervals:
$$loss(x,y) = \frac{|upper(\mathcal{V}(x,y) - lower(\mathcal{V}(x,y)))|}{|max(x) - min(x)|}$$

where $leafs(x,v)$ is the number of leaf nodes at the value $v$ in the hierarchy of attribute $x$, $\mathcal{V}(x,y)$ is the value of attribute $x$ in record $y$, $upper(v)$ and $lower(v)$ are the upper and lower bounds of an interval $v$ respectively. If all values $\mathcal{V}(x,y)$ are suppressed, then $loss(x,y)$ = 1. We select the solution with high utility.

We considered two privacy models ($k$-anonymity, $\ell$-diversity) for all our experiments. As defined in Definition \ref{def:tcloseness}, $t$-closeness caters for the distribution of sensitive attributes which does not affect the results of using machine learning (ML) models. Therefore, we exclude this privacy model from our analysis. For the ML model, we used a 3-layer multilayer perceptron (MLP).

\begin{figure}[!tpb]
\centerline{\includegraphics[width=\linewidth]{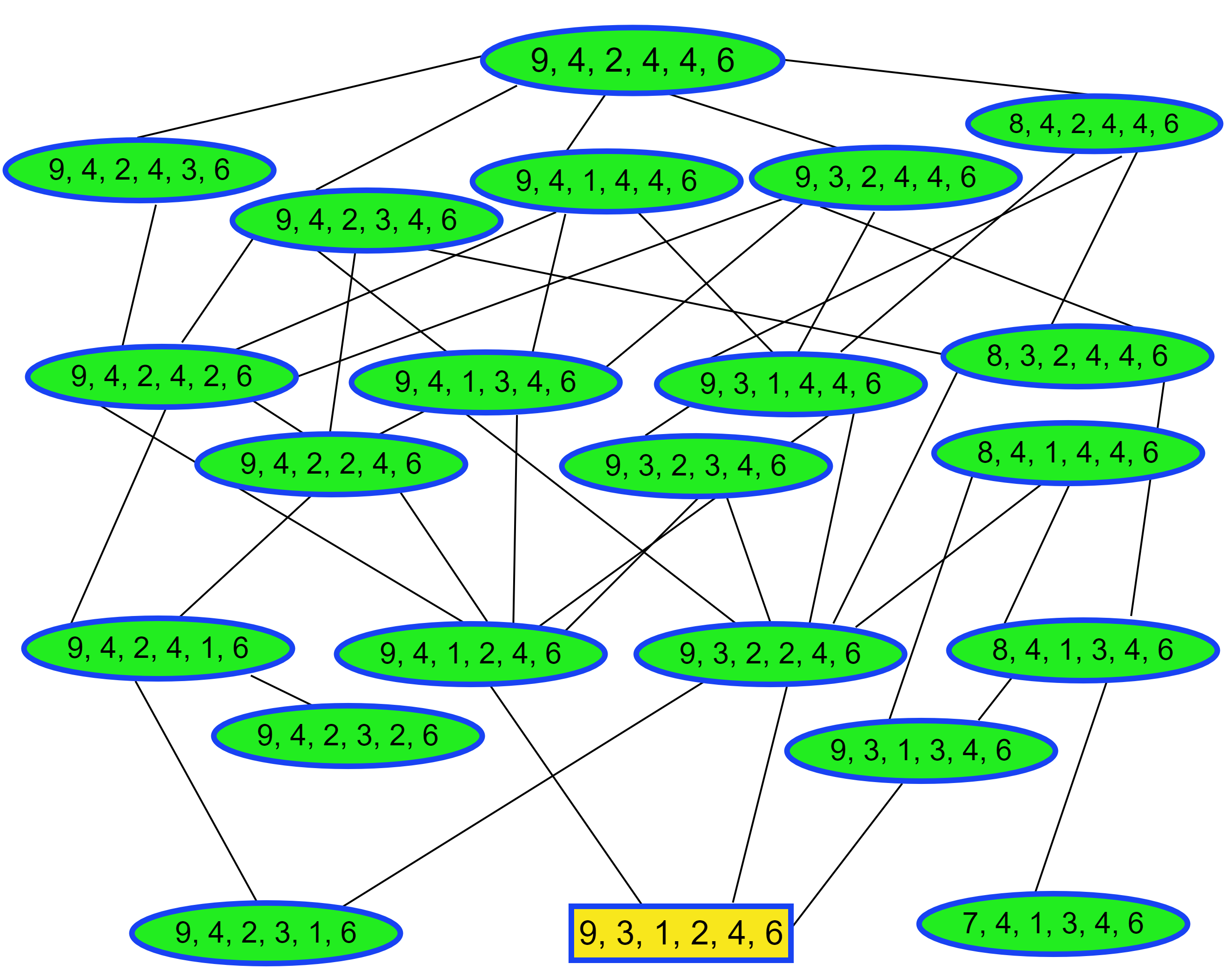}}
\caption{Solution space lattice of performing 100-anonymity when all demographics are anonymized. The highlighted yellow node shows the optimum solution. Different numbers indicate the height of the hierarchy tree of each demographic attribute that satisfies the privacy model.}\label{solutionspacelattice}
\end{figure}

\begin{table}[!t]
\processtable{Performance of using other features to predict demographics (reconstruction attack).\label{Tab:demopred}} 
{\begin{tabular}{@{}ll@{}}\toprule Feature &
Accuracy \\\midrule
Age & $0.77 \pm 0.006$\\
Ethnicity & $ 0.73 \pm 0.004 $\\
Marital Status & $ 0.50 \pm 0.005 $\\
Insurance & $ 0.58 \pm 0.002 $\\
Religion & $ 0.38 \pm 0.003 $\\
Gender & $ 0.57 \pm 0.001 $\\\botrule
\end{tabular}}{}
\end{table}

\end{methods}

\subsection{Results and discussion}
\label{sec:expresults}

\begin{figure}[!tpb]
\centering
\begin{subfigure}[b]{.45\linewidth}
\centering
\includegraphics[width=\linewidth]{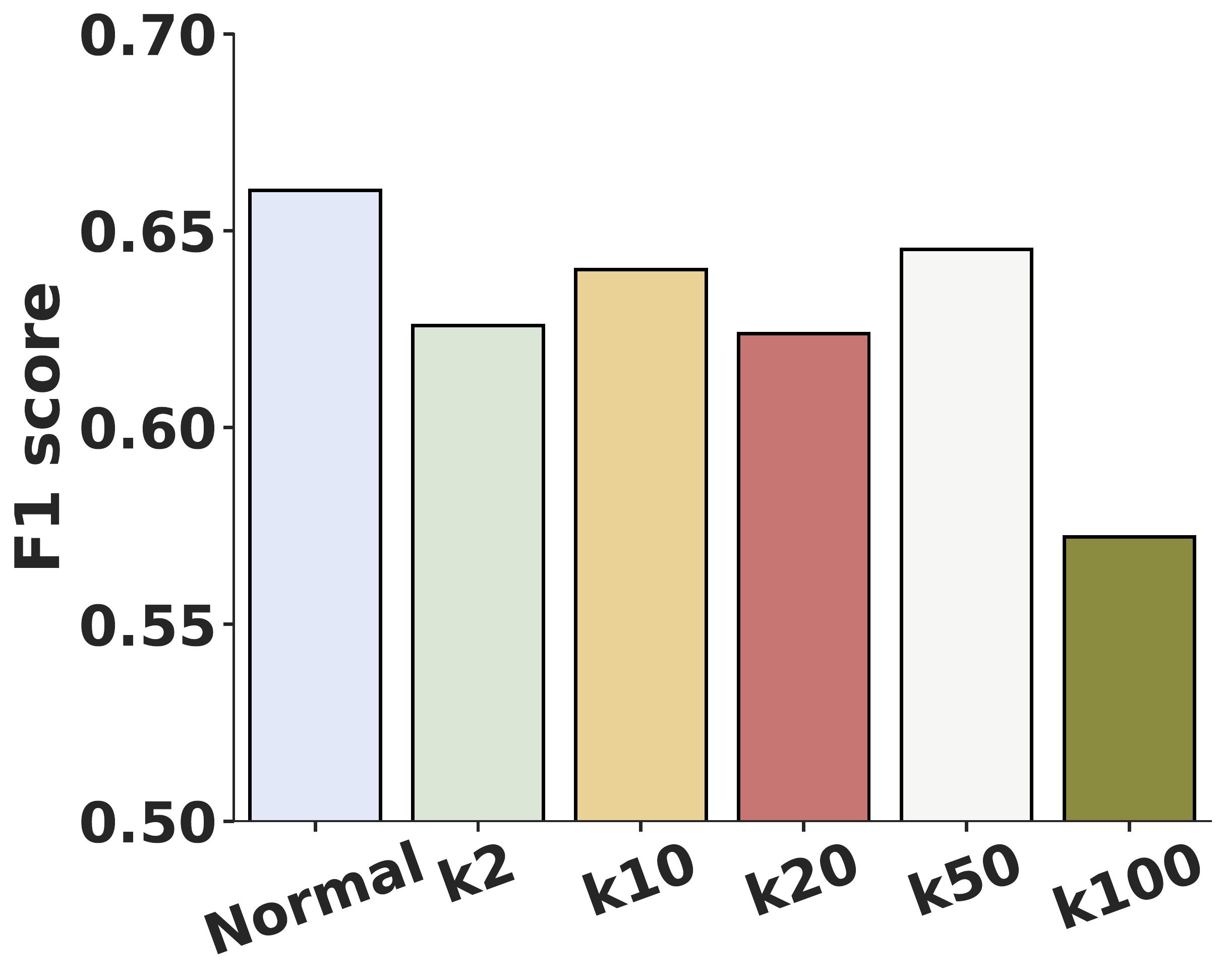}
\caption{24hrs Readmission}
\label{sfig:24hrsReadmission}
\end{subfigure}%
\begin{subfigure}[b]{.45\linewidth}
\centering
\includegraphics[width=\linewidth]{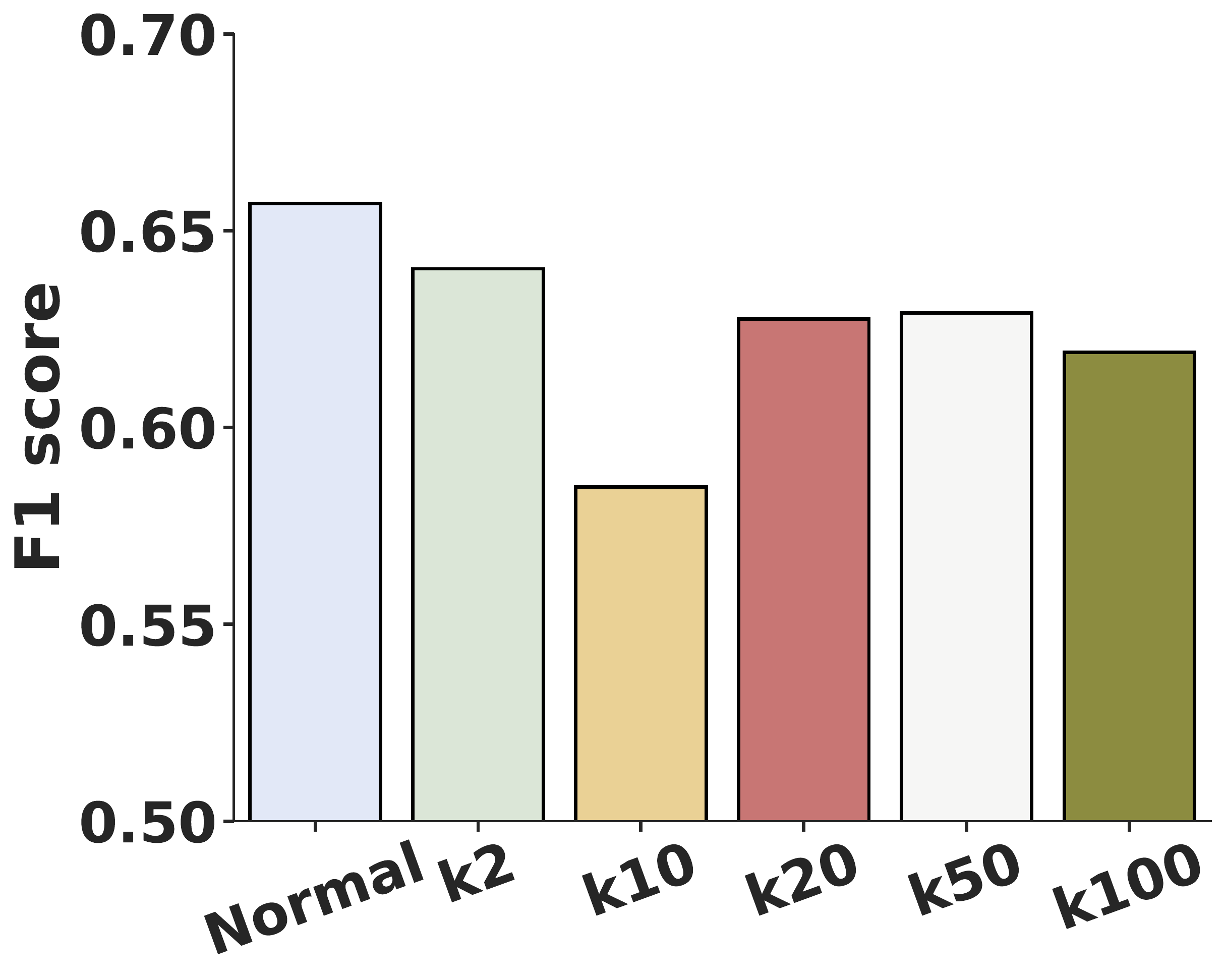}
\caption{48hrs Readmission}
\label{sfig:48hrsReadmission}
\end{subfigure}%

\begin{subfigure}[b]{.45\linewidth}
\centering
\includegraphics[width=\linewidth]{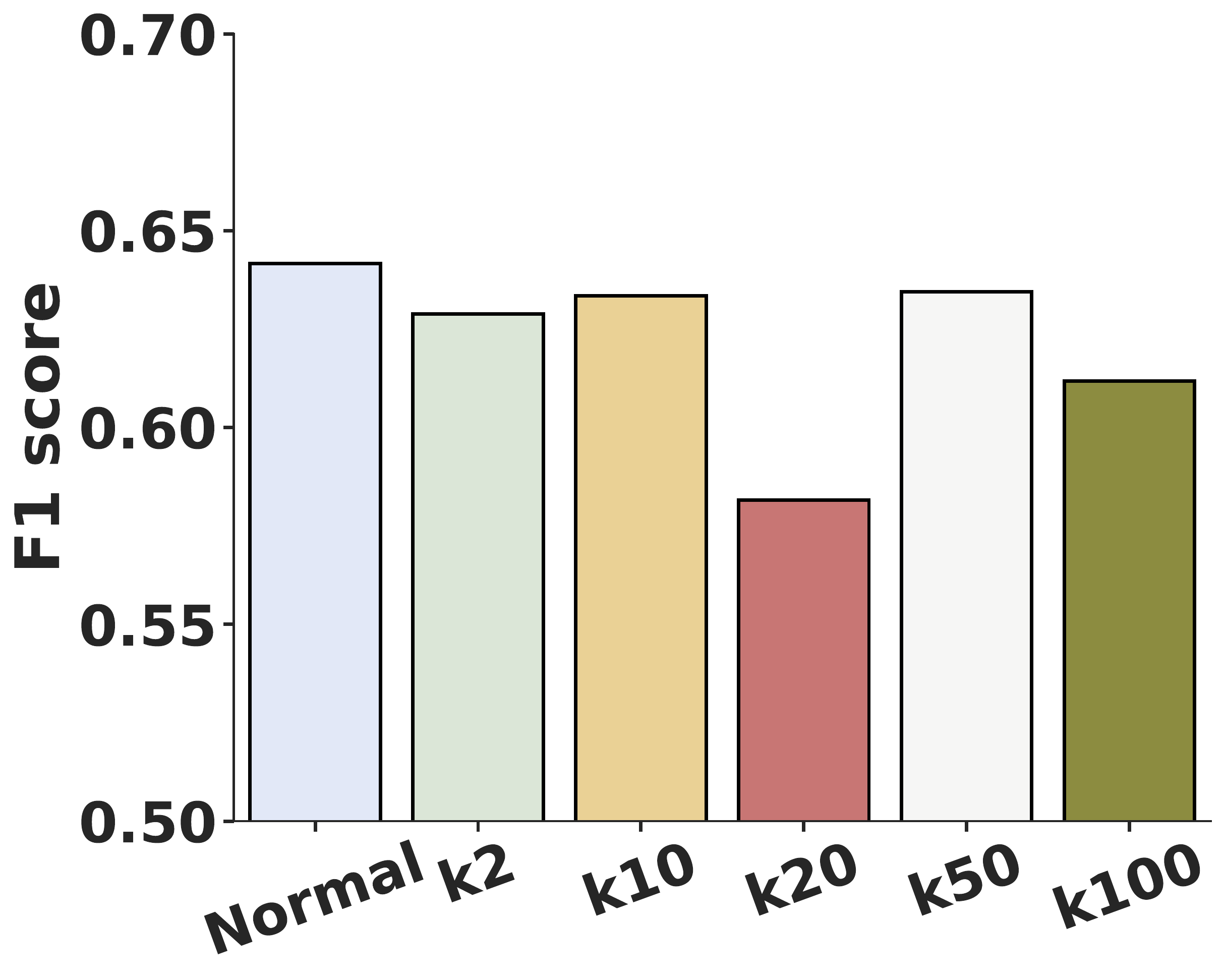}
\caption{72hrs Readmission}
\label{sfig:72hrsReadmission}
\end{subfigure}%
\begin{subfigure}[b]{.45\linewidth}
\centering
\includegraphics[width=\linewidth]{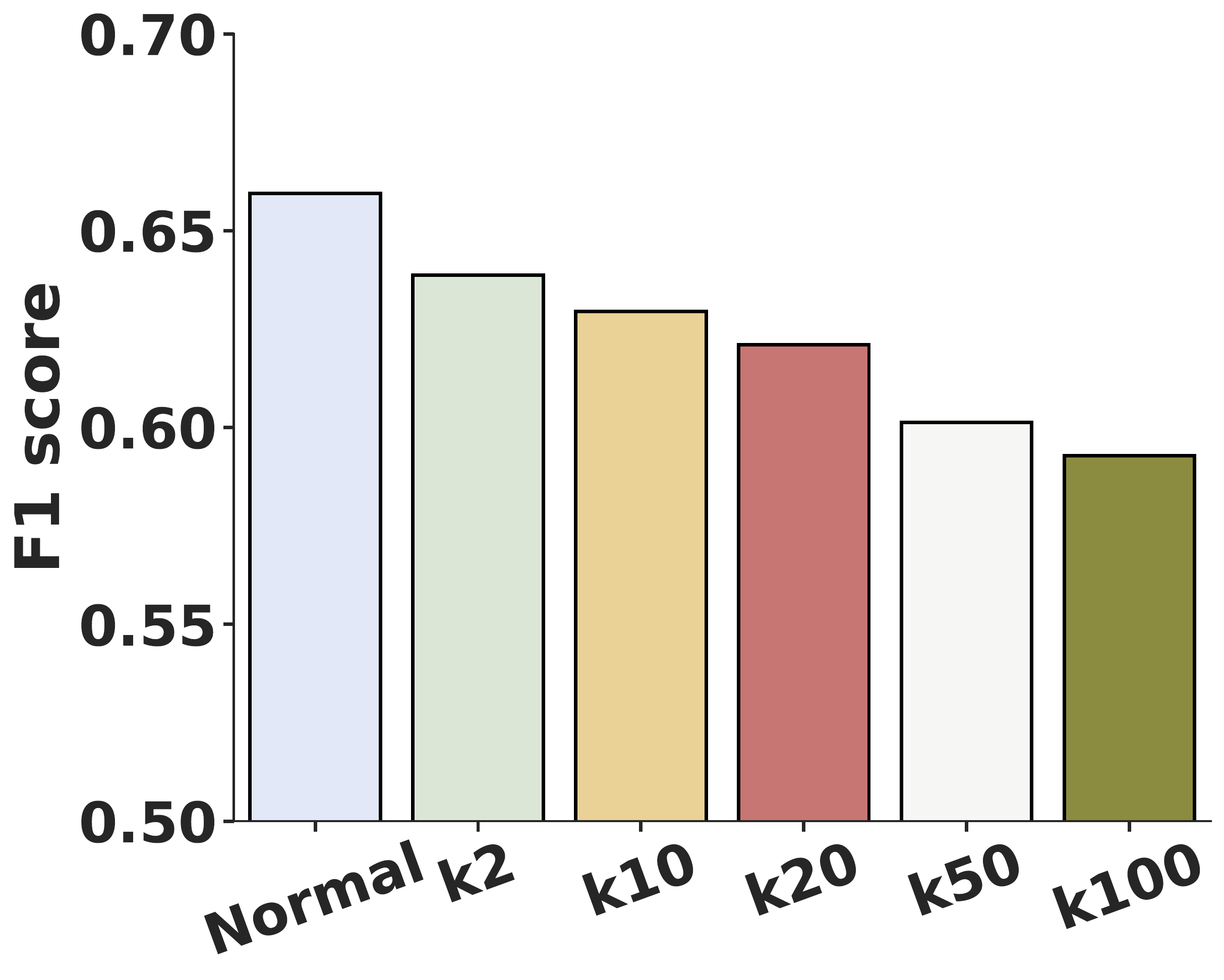}
\caption{7days Readmission}
\label{sfig:7daysReadmission}
\end{subfigure}%

\begin{subfigure}[b]{.45\linewidth}
\centering
\includegraphics[width=\linewidth]{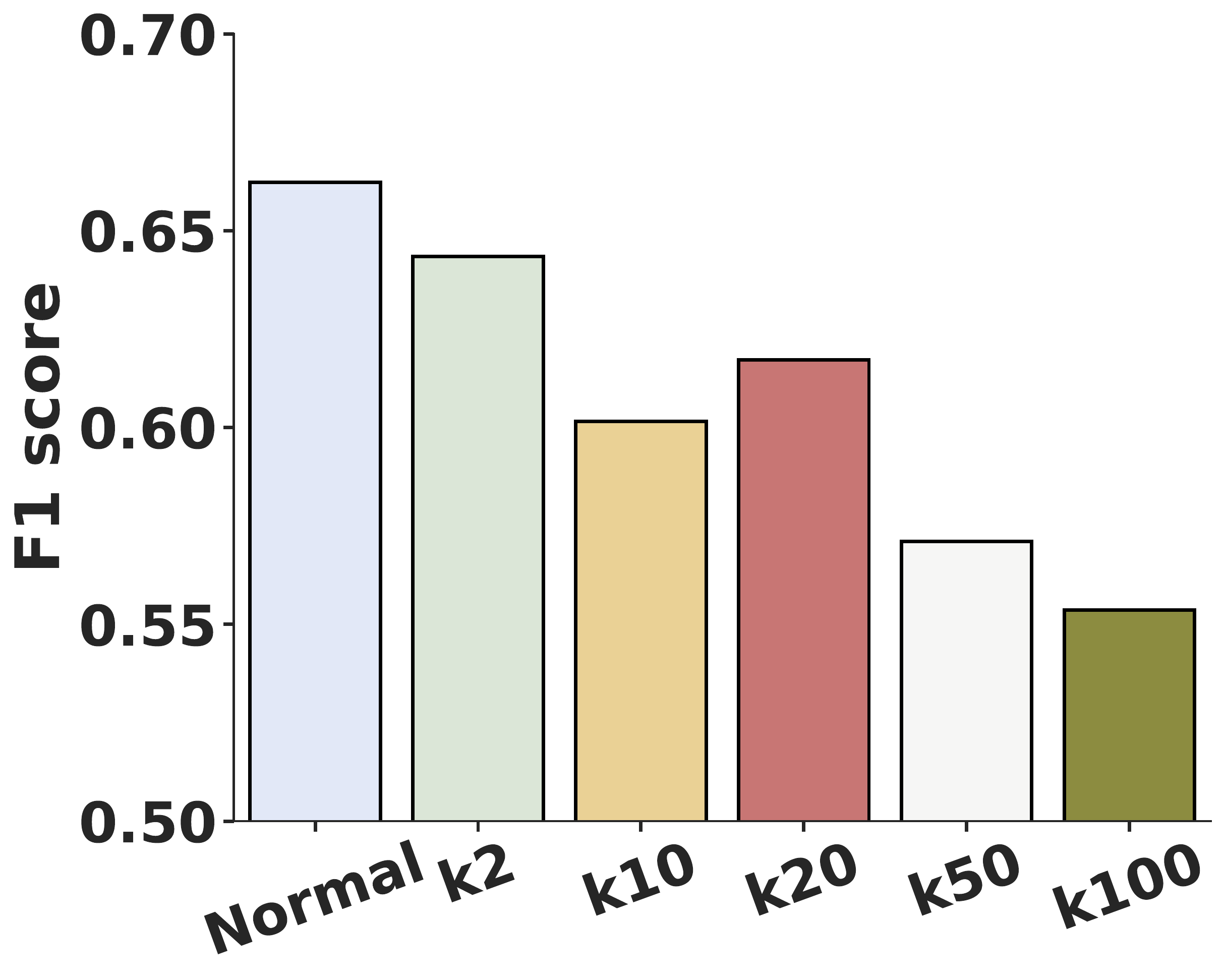}
\caption{30days Readmission}
\label{sfig:30daysReadmission}
\end{subfigure}%
\begin{subfigure}[b]{.45\linewidth}
\centering
\includegraphics[width=\linewidth]{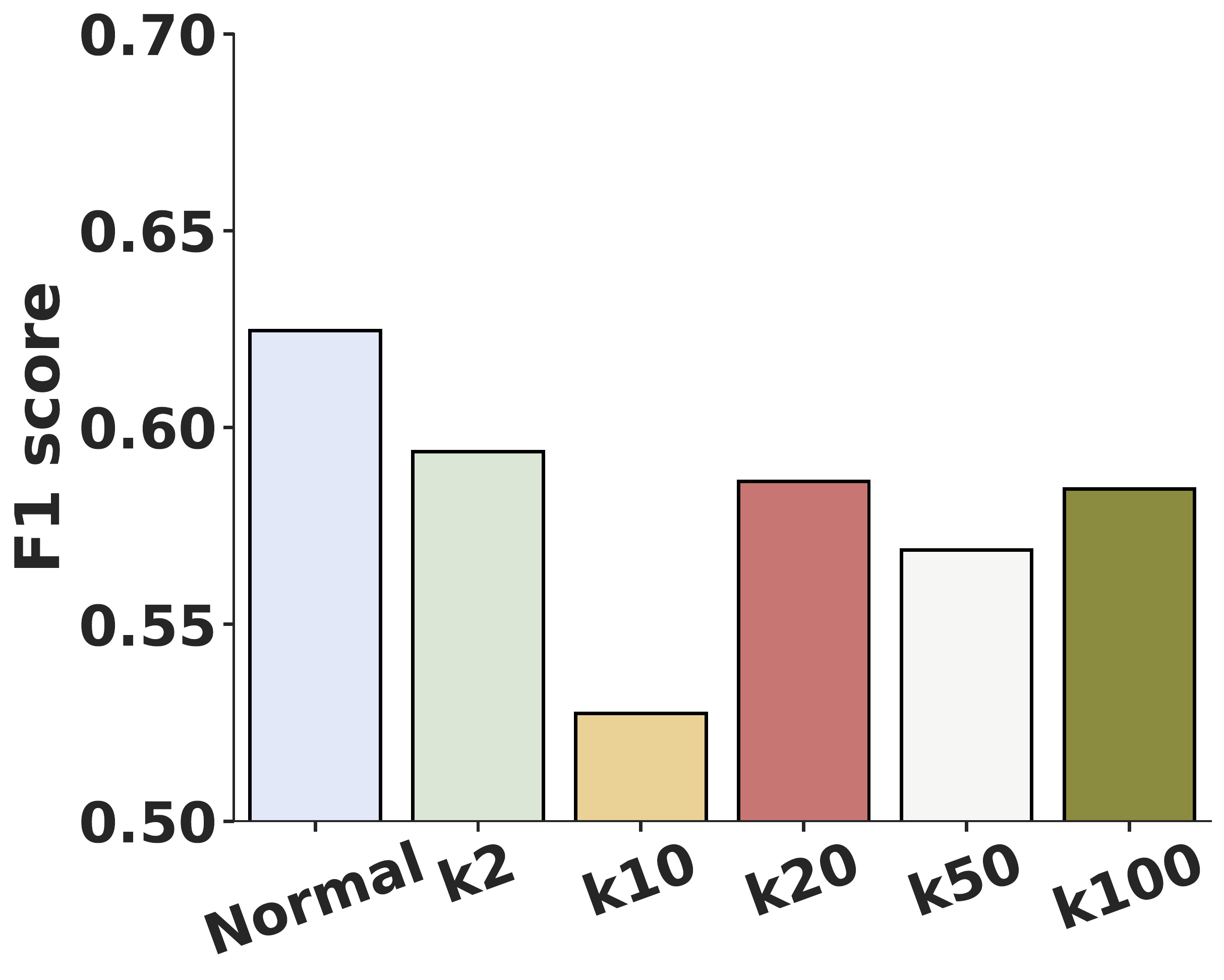}
\caption{Rebounce}
\label{sfig:Rebounce}
\end{subfigure}%

\caption{Performance of training the ML model for patient's readmission prediction task using only demographic data.}
\label{readmission}
\end{figure}

\subsubsection{Reconstruction attack: result}
Following the observation from \citet{tang2018predictive} where demographic information does not affect the risk of readmission prediction. We reverse the process by first using only the demographics features to predict the risk of readmission. Specifically, we ignore the $x\_19$ features. As shown in Figure \ref{readmission}, demographic features can still predict with some reasonable performance. This implies that there is some correlation between $x\_19$ and demographics features. Therefore, the goal of our reconstruction attack is to construct the non-sanitized demographic features using the anonymized $x\_19$ features.
Table \ref{Tab:demopred} shows the accuracy of the ML model for our reconstruction attack. This shows that with high accuracy, demographics such as age, ethnicity, insurance, and gender can be reconstructed. However, marital status and religion are difficult to reconstruct because these attributes have been over-generalized by the privacy models.

\subsubsection{Effect of generalization and suppression}
\begin{figure}[!tpb]
\centerline{\includegraphics[width=\linewidth]{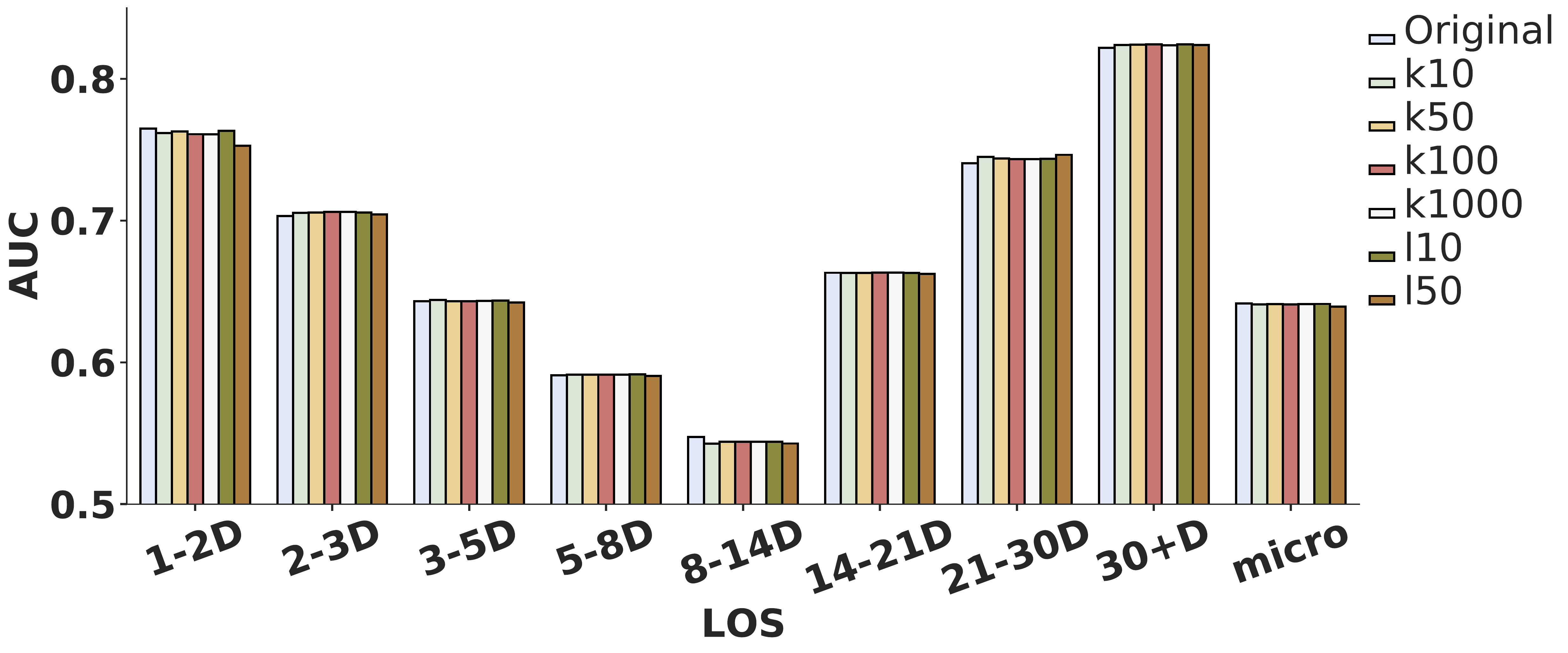}}
\caption{Performance of the ML model when temperature and glucose are the anonymized attributes on the LOS prediction task.}\label{tempglu}
\end{figure}

\begin{figure}[!tpb]
\centerline{\includegraphics[width=\linewidth]{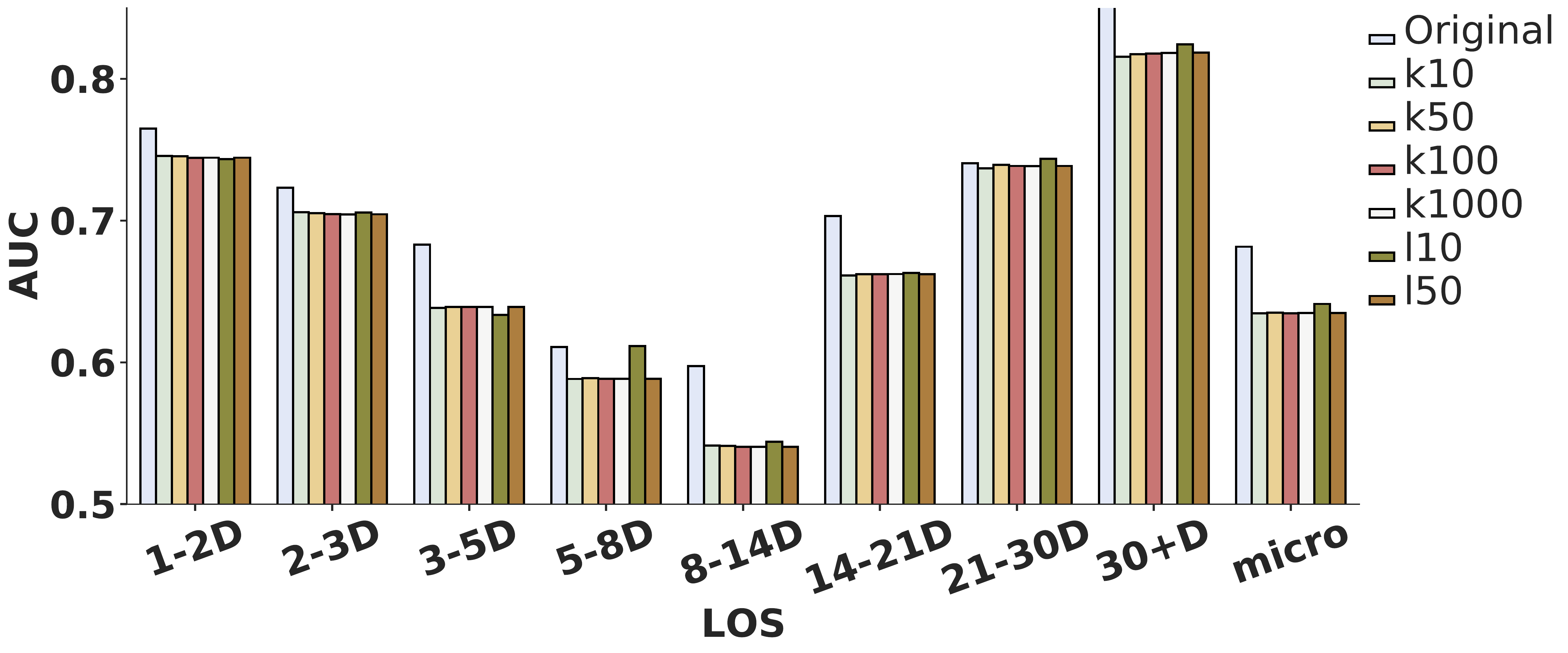}}
\caption{Performance of the ML model when only the heart rate measurement attribute is anonymized on the LOS prediction task.}\label{heart}
\end{figure}

\begin{figure}[!tpb]
\centerline{\includegraphics[width=\linewidth]{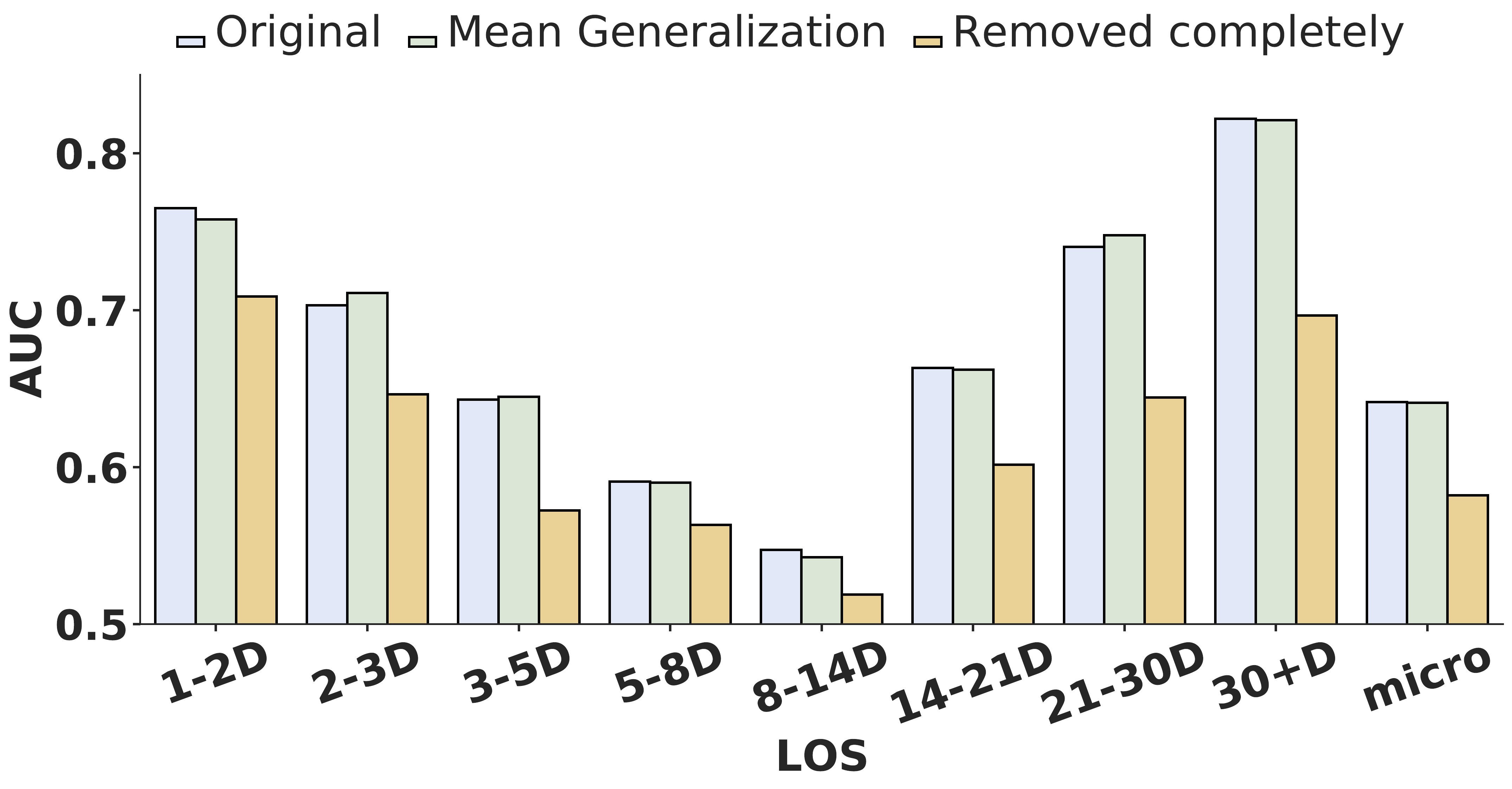}}
\caption{The effect of mean generalization and suppression of data that does not meet privacy model constraints for $k$-anonymity where k=1000.}\label{k1kcompleteremove}
\end{figure}

To quantify the effect of generalization and suppression on the privacy model, we used the length of stay (LOS) prediction task. We first anonymize the original data according to the privacy model, then run the ML model on the anonymized data.
As shown in Figure \ref{k1kcompleteremove}, when all $x\_19$ features are anonymized by mean generalization, the performance of the anonymized data and the non-sanitized data does not differ significantly. However, when the data records that violate the privacy model constraint are completely removed (suppression), the performance drops significantly. To mitigate this effect, we turn to the $LKC$-privacy model \cite{cite732mohammed2010centralized}. The $LKC$-privacy model \cite{cite732mohammed2010centralized} requires that not all attributes need to be anonymized but a combination of $L$ attributes. However, since combinatorics suffers from the curse of dimensionality, we adopted a feature selection method on the non-sanitized data for ranking the features to determine the predominant ones. We then anonymize the predominant features rather than anonymizing all $x\_19$ features. According to the feature ranking, heart rate measurement is the top feature while temperature and glucose are the least features. Figure \ref{tempglu} shows the effect of anonymizing temperature and glucose. We observe that the anonymized data does not differ much from the non-sanitized data across all privacy models. This is because the feature ranking of these attributes is low (bottom 2). As shown in Figure \ref{heart}, when the top feature is anonymized, the difference between the anonymized data and non-sanitized data is significant. These observations show that attributes with the least effect on prediction performance are indifferent when they are anonymized. Similarly, anonymizing only the top features have a major effect on the performance.
Therefore attribute level privacy based on $LKC$-privacy model is better than anonymizing all attributes provided that the anonymized attribute is the knowledge that the adversary can have.

\section{Differential privacy as a defense mechanism}
\label{sec:soaprivacypreservetech}
To defend against attacks on anonymized data as discussed in Sections \ref{sec:graphattacks} and \ref{sec:structuredattacks}, differential privacy (DP) was proposed \cite{CynthiaDwork10.1561/0400000042}. 
DP is a mathematical definition of privacy aimed at preserving the privacy of an individual via the addition of noise. The main intuition of DP is a guarantee that the inclusion or exclusion of an individual's record has little effect on the output of the analysis. Consider two datasets $D$, $D'$ where they differ in at most one data record (neighbors). DP ensures that the output of performing analysis on $D$ and $D'$ will be the same. This guarantees that no individual record can be inferred. Thus, robust to reconstruction attack.

Noise addition methods to satisfy DP can be local or global.
In \textit{local DP}, noise is added to each data point in the dataset (either by a dataset curator once the dataset is formed or by the individuals themselves before making their data available to the curator) whereas in \textit{global DP}, the noise necessary to protect the individual's privacy is added at the output of the query of the dataset. Generally, global DP can lead to more accurate results when compared to local DP while keeping the same privacy level. However, when using global differential privacy, data owners need to trust the dataset curator to add the necessary noise to preserve their privacy.

\mpara{DP for answering queries.} DP has been applied to privacy-preserving data mining \cite{cite412friedman2010data, cite413soria2017individual, abadi2016deep, cite414sarathy2011evaluating, cite415mcsherry2009privacy, cite416roy2010airavat, cite417mohan2012gupt} where the goal is to ensure indistinguishability of results between any pair of neighbor data sets which differs by one record by adding noise.  An alternative approach to standard DP based on individual differential privacy was proposed in \cite{cite413soria2017individual} where utility is preserved by introducing less noise to the query result. This is achieved by assuming that the data curator can use her knowledge of the actual data set at the time of query response and downwardly adjust the distortion to the actual data. \citet{cite414sarathy2011evaluating} evaluated the effect of adding Laplace noise to queries to ensure privacy. They showed that the addition of Laplace noise to queries (since it depends on the sensitivity of the query function and privacy budget) can still be vulnerable to tracker attack where an adversary can issue multiple queries and use the query response to uncover one or more observation. In order to avoid such vulnerability, the noise variance added should increase as a function of the number of queries.

Platforms that ensure DP includes privacy-integrated queries (PINQ) \cite{cite415mcsherry2009privacy} that provides SQL-like language to analysts while protecting privacy and GUPT \cite{cite417mohan2012gupt} which ensures that the privacy budget is distributed among queries based on the desired level of privacy and utility.

DP is a natural fit for query processing rather than data publishing since the goal of DP is to ensure that the output of two neighboring datasets is the same. However, applying DP to data publishing is desirable due to its guarantees.

\mpara{DP for data publishing of health microdata.} The application of DP to data publishing is mainly by releasing aggregated results based on count queries or contingency tables or histograms \cite{hardt2010simple, li2015differentially}. For example, a contingency table can be created by combining demographics and LOS where LOS is the sensitive attribute. To achieve DP, noise is added to the counts of the different combinations of demographic attributes and LOS. However, the more the attributes, the more the noise that needs to be added to ensure DP which leads to higher data distortion. Similarly, creating a contingency table for health data with several attributes is difficult in practice. Moreover, since ML models aim at learning the dependencies between several attributes to make predictions, releasing microdata is preferred over contingency tables.
\citet{soria2014enhancing} proposed an intuitive approach to achieving DP based on $k$-anonymity. Their method is based on microaggregation technique that adds noise to the $k$-anonymous version of the data set. \citet{zhang2017privbayes} proposed a deferentially private method for releasing high-dimensional data by constructing a Bayesian network. The Bayesian network models the correlations among the attributes and approximate the distribution of the data using a set of low-dimensional marginals. Then noise is added to each marginal to ensure DP. Both the noisy marginals and the Bayesian network are then used to construct an approximation of the data distribution. Instead of releasing the dataset of the approximate distribution, they first sample records from the approximate distribution to construct a synthetic dataset, then releases the synthetic data.

A recent approach to releasing microdata for health data employed data perturbation techniques to reduce the amount of noise required to satisfy DP \cite{lee2020differentially}. Specifically, generalization, suppression, and insertion of random data (noisy insertion) are first performed on the raw data. The generalization and suppression reduce the amount of noise needed to be added because the combination of attributes will be reduced. Then a utility score based on information loss is assigned to each record. Finally, records with less information loss are selected as candidates to be released. The noisy insertion ensures that the released data satisfies DP and thus robust to reconstruction attack.

\section{Tools}
\label{sec:tools}
\mpara{ARX \cite{prasser2020flexible}.} ARX is a non-interactive microdata anonymization tool that tends to automate the anonymization process. This means that protected records are created from records of an input dataset. This is also known as one-time anonymization or release and forget where the publisher anonymizes the database and then publishes it, allowing third parties to access the anonymized data. It combines several transformation techniques such as sampling, aggregation, suppression, categorization, and generalization. The ARX data anonymization tool is an open software that supports an arbitrary combination of privacy and utility models. Thus, making it a generic anonymization tool. Privacy models supported include $k$-anonymity, $\ell$-diversity, $t$-closeness, $\delta$-disclosure privacy, $\beta$-likeness and $\delta$-presence, $k$-map and DP.

\mpara{PySyft \cite{ryffel2018genericPysyft}.} PySyft is an extension built on top of popular deep learning frameworks such as PyTorch and Tensorflow for encrypted privacy-preserving deep learning. It is a Python library for secure and private deep learning. It supports federated learning, DP, and encrypted computation including multi-party computation and homomorphic encryption. Although PySyft is in its early years of development, the scalability and privacy protection it offers is very promising.

\mpara{Anonimatron \cite{anonimatron}.} Anonimatron is an open-source data anonymization tool written in Java. It can generate surrogate data with the same properties as the original data. It can also generate fake email addresses, fake Roman names, and universally unique identifiers (UUIDs). Anonimatron supports popular DBMS including Oracle, PostgreSQL, and MySQL.

\mpara{Synthea \cite{Synthea10.1093/jamia/ocx079}.} Synthea is an open-source synthetic medical data generation tool that models the medical history of patients for research. It utilizes realistic but not real medical data. Synthea uses the PADARSER framework \cite{dube2013approach} to generate a skeletal synthetic EHR. As output, Synthea provides fast healthcare interoperability resources(FHIR), a medical information standard for exchanging EHR. Synthea generates realistic medical data for patients living in the imaginary commonwealth of Massachusetts. A complete lifetime EHR was obtained for each patient from birth to death using different publicly available sources like US Census Bureau demographics or National Institutes of Health reports.

Other tools include UTD Anonymization Toolbox, Cornell Anonymization Toolkit, TIA-MAT, Anamnesia, SECRETA, sdcMicro and $\mu$-Argus. However, they usually only support a limited set of privacy models or focus on specific privacy and data transformation models.

\section{Conclusion}
In this paper, we provided a comprehensive review of anonymization models and techniques applicable for relational and graph-based health care data. Besides, we studied possible attacks on anonymized data and empirically demonstrated reconstruction attack on MIMIC-III data. Finally we discussed existing defense mechanisms while giving an overview of existing anonymization tools. We believe that our comprehensive review covering different perspectives on anonymization will assist researchers and practitioners in selecting relevant anonymization techniques based on the data type, desired privacy level, information loss, and possible adversarial behavior.





\section*{Funding}
This work is in part funded by the Lower Saxony Ministry of Science and Culture under grant number ZN3491 within the Lower Saxony "Vorab" of the Volkswagen Foundation and supported by the Center for Digital Innovations (ZDIN), and the Federal Ministry of Education and Research (BMBF), Germany under the project LeibnizKILabor (grant number 01DD20003).

\vspace*{-12pt}
\bibliographystyle{plainnat}
\bibliography{main}

\begin{thebibliography}{118}
\providecommand{\natexlab}[1]{#1}
\providecommand{\url}[1]{\texttt{#1}}
\expandafter\ifx\csname urlstyle\endcsname\relax
  \providecommand{\doi}[1]{doi: #1}\else
  \providecommand{\doi}{doi: \begingroup \urlstyle{rm}\Url}\fi

\bibitem[Abadi et~al.(2016)Abadi, Chu, Goodfellow, McMahan, Mironov, Talwar,
  and Zhang]{abadi2016deep}
Martin Abadi, Andy Chu, Ian Goodfellow, H~Brendan McMahan, Ilya Mironov, Kunal
  Talwar, and Li~Zhang.
\newblock Deep learning with differential privacy.
\newblock In \emph{Proceedings of the 2016 ACM SIGSAC Conference on Computer
  and Communications Security}, pages 308--318, 2016.

\bibitem[Agarwal and Sachdeva(2018)]{agarwal2018enhanced}
Shivam Agarwal and Shelly Sachdeva.
\newblock An enhanced method for privacy-preserving data publishing.
\newblock In \emph{Innovations in Computational Intelligence}, pages 61--75.
  Springer, 2018.

\bibitem[Aggarwal(2005)]{cite2aggarwal2005k}
Charu~C Aggarwal.
\newblock On k-anonymity and the curse of dimensionality.
\newblock In \emph{VLDB}, volume~5, pages 901--909, 2005.

\bibitem[Aggarwal(2008)]{cite723aggarwal2008privacy}
Charu~C Aggarwal.
\newblock Privacy and the dimensionality curse.
\newblock In \emph{Privacy-Preserving Data Mining}, pages 433--460. Springer,
  2008.

\bibitem[Agrawal et~al.(2003)Agrawal, Evfimievski, and
  Srikant]{cite63agrawal2003information}
Rakesh Agrawal, Alexandre Evfimievski, and Ramakrishnan Srikant.
\newblock Information sharing across private databases.
\newblock In \emph{Proceedings of the 2003 ACM SIGMOD international conference
  on Management of data}, pages 86--97, 2003.

\bibitem[Andreou et~al.(2017)Andreou, Goga, and
  Loiseau]{cite5andreou2017identity}
Athanasios Andreou, Oana Goga, and Patrick Loiseau.
\newblock Identity vs. attribute disclosure risks for users with multiple
  social profiles.
\newblock In \emph{Proceedings of the 2017 IEEE/ACM International Conference on
  Advances in Social Networks Analysis and Mining 2017}, pages 163--170, 2017.

\bibitem[Anonimatron(2015)]{anonimatron}
Anonimatron.
\newblock Gdpr compliant testing., 2015.
\newblock URL \url{https://realrolfje.github.io/anonimatron/}.

\bibitem[Bearman et~al.(2004)Bearman, Moody, and Stovel]{bearman2004chains}
Peter~S Bearman, James Moody, and Katherine Stovel.
\newblock Chains of affection: The structure of adolescent romantic and sexual
  networks.
\newblock \emph{American journal of sociology}, 110\penalty0 (1):\penalty0
  44--91, 2004.

\bibitem[Bhagat et~al.(2009)Bhagat, Cormode, Krishnamurthy, and
  Srivastava]{bhagat2009class}
Smriti Bhagat, Graham Cormode, Balachander Krishnamurthy, and Divesh
  Srivastava.
\newblock Class-based graph anonymization for social network data.
\newblock \emph{Proceedings of the VLDB Endowment}, 2\penalty0 (1):\penalty0
  766--777, 2009.

\bibitem[Branting et~al.(2016)Branting, Reeder, Gold, and
  Champney]{branting2016graph}
L~Karl Branting, Flo Reeder, Jeffrey Gold, and Timothy Champney.
\newblock Graph analytics for healthcare fraud risk estimation.
\newblock In \emph{2016 IEEE/ACM International Conference on Advances in Social
  Networks Analysis and Mining (ASONAM)}, pages 845--851. IEEE, 2016.

\bibitem[Budiardjo et~al.(2019{\natexlab{a}})Budiardjo, Wibowo, Achsan,
  et~al.]{budiardjo2019approach}
Eko~K Budiardjo, Wahyu~C Wibowo, Harry~TY Achsan, et~al.
\newblock An approach for distributing sensitive values in k-anonymity.
\newblock In \emph{2019 International Workshop on Big Data and Information
  Security (IWBIS)}, pages 109--114. IEEE, 2019{\natexlab{a}}.

\bibitem[Budiardjo et~al.(2019{\natexlab{b}})Budiardjo, Wibowo,
  et~al.]{budiardjo2019privacy}
Eko~Kuswardono Budiardjo, Wahyu~Catur Wibowo, et~al.
\newblock Privacy preserving data publishing with multiple sensitive attributes
  based on overlapped slicing.
\newblock \emph{Information}, 10\penalty0 (12):\penalty0 362,
  2019{\natexlab{b}}.

\bibitem[Byun et~al.(2006)Byun, Sohn, Bertino, and Li]{cite724byun2006secure}
Ji-Won Byun, Yonglak Sohn, Elisa Bertino, and Ninghui Li.
\newblock Secure anonymization for incremental datasets.
\newblock In \emph{Workshop on secure data management}, pages 48--63. Springer,
  2006.

\bibitem[Byun et~al.(2007)Byun, Kamra, Bertino, and Li]{cite6byun2007efficient}
Ji-Won Byun, Ashish Kamra, Elisa Bertino, and Ninghui Li.
\newblock Efficient k-anonymization using clustering techniques.
\newblock In \emph{International Conference on Database Systems for Advanced
  Applications}, pages 188--200. Springer, 2007.

\bibitem[Campan et~al.(2010)Campan, Truta, and Cooper]{campan2010p}
Alina Campan, Traian~Marius Truta, and Nicholas Cooper.
\newblock P-sensitive k-anonymity with generalization constraints.
\newblock \emph{Trans. Data Priv.}, 3\penalty0 (2):\penalty0 65--89, 2010.

\bibitem[Cao et~al.(2011)Cao, Karras, Kalnis, and Tan]{cao2011sabre}
Jianneng Cao, Panagiotis Karras, Panos Kalnis, and Kian-Lee Tan.
\newblock Sabre: a sensitive attribute bucketization and redistribution
  framework for t-closeness.
\newblock \emph{The VLDB Journal}, 20\penalty0 (1):\penalty0 59--81, 2011.

\bibitem[Cheng et~al.(2010)Cheng, Fu, and Liu]{cite5cheng2010k}
James Cheng, Ada Wai-chee Fu, and Jia Liu.
\newblock K-isomorphism: privacy preserving network publication against
  structural attacks.
\newblock In \emph{Proceedings of the 2010 ACM SIGMOD International Conference
  on Management of data}, pages 459--470, 2010.

\bibitem[Cong et~al.(2018)Cong, Feng, Li, Zhang, Rao, and
  Tao]{cong2018constructing}
Qing Cong, Zhiyong Feng, Fang Li, Li~Zhang, Guozheng Rao, and Cui Tao.
\newblock Constructing biomedical knowledge graph based on semmeddb and linked
  open data.
\newblock In \emph{2018 IEEE International Conference on Bioinformatics and
  Biomedicine (BIBM)}, pages 1628--1631. IEEE, 2018.

\bibitem[Domingo-Ferrer and Torra(2005)]{cite7domingo2005ordinal}
Josep Domingo-Ferrer and Vicen{\c{c}} Torra.
\newblock Ordinal, continuous and heterogeneous k-anonymity through
  microaggregation.
\newblock \emph{Data Mining and Knowledge Discovery}, 11\penalty0 (2):\penalty0
  195--212, 2005.

\bibitem[Dong et~al.(2005)Dong, Halevy, and Madhavan]{cite68dong2005reference}
Xin Dong, Alon Halevy, and Jayant Madhavan.
\newblock Reference reconciliation in complex information spaces.
\newblock In \emph{Proceedings of the 2005 ACM SIGMOD international conference
  on Management of data}, pages 85--96, 2005.

\bibitem[Du et~al.(2008)Du, Teng, and Zhu]{cite436du2008privacy}
Wenliang Du, Zhouxuan Teng, and Zutao Zhu.
\newblock Privacy-maxent: integrating background knowledge in privacy
  quantification.
\newblock In \emph{Proceedings of the 2008 ACM SIGMOD international conference
  on Management of data}, pages 459--472, 2008.

\bibitem[Dube and Gallagher(2013)]{dube2013approach}
Kudakwashe Dube and Thomas Gallagher.
\newblock Approach and method for generating realistic synthetic electronic
  healthcare records for secondary use.
\newblock In \emph{International Symposium on Foundations of Health Informatics
  Engineering and Systems}, pages 69--86. Springer, 2013.

\bibitem[Dwork and Roth(2014)]{CynthiaDwork10.1561/0400000042}
Cynthia Dwork and Aaron Roth.
\newblock The algorithmic foundations of differential privacy.
\newblock \emph{Found. Trends Theor. Comput. Sci.}, 9\penalty0
  (3–4):\penalty0 211–407, August 2014.
\newblock ISSN 1551-305X.
\newblock \doi{10.1561/0400000042}.
\newblock URL \url{https://doi.org/10.1561/0400000042}.

\bibitem[El~Emam and Dankar(2008)]{cite725el2008protecting}
Khaled El~Emam and Fida~Kamal Dankar.
\newblock Protecting privacy using k-anonymity.
\newblock \emph{Journal of the American Medical Informatics Association},
  15\penalty0 (5):\penalty0 627--637, 2008.

\bibitem[Eze and Peyton(2015)]{cite7eze2015systematic}
Benjamin Eze and Liam Peyton.
\newblock Systematic literature review on the anonymization of high dimensional
  streaming datasets for health data sharing.
\newblock \emph{Procedia Computer Science}, 63:\penalty0 348--355, 2015.

\bibitem[Foffano et~al.(2019)Foffano, Rossi, and Torsello]{foffano2019you}
Daniele Foffano, Luca Rossi, and Andrea Torsello.
\newblock You can't see me: Anonymizing graphs using the szemer{\'e}di
  regularity lemma.
\newblock \emph{Frontiers in Big Data}, 2:\penalty0 7, 2019.

\bibitem[Friedman and Schuster(2010)]{cite412friedman2010data}
Arik Friedman and Assaf Schuster.
\newblock Data mining with differential privacy.
\newblock In \emph{Proceedings of the 16th ACM SIGKDD international conference
  on Knowledge discovery and data mining}, pages 493--502, 2010.

\bibitem[Fung et~al.(2007)Fung, Wang, and Philip]{cite442fung2007anonymizing}
Benjamin~CM Fung, Ke~Wang, and S~Yu Philip.
\newblock Anonymizing classification data for privacy preservation.
\newblock \emph{IEEE transactions on knowledge and data engineering},
  19\penalty0 (5):\penalty0 711--725, 2007.

\bibitem[Fung et~al.(2010)Fung, Wang, Fu, and
  Philip]{cite211fung2010introduction}
Benjamin~CM Fung, Ke~Wang, Ada Wai-Chee Fu, and S~Yu Philip.
\newblock \emph{Introduction to privacy-preserving data publishing: Concepts
  and techniques}.
\newblock CRC Press, 2010.

\bibitem[Fung et~al.(2011)Fung, Trojer, Hung, Xiong, Al-Hussaeni, and
  Dssouli]{cite726fung2011service}
Benjamin~CM Fung, Thomas Trojer, Patrick~CK Hung, Li~Xiong, Khalil Al-Hussaeni,
  and Rachida Dssouli.
\newblock Service-oriented architecture for high-dimensional private data
  mashup.
\newblock \emph{IEEE Transactions on Services Computing}, 5\penalty0
  (3):\penalty0 373--386, 2011.

\bibitem[Gal et~al.(2014)Gal, Tucker, Gangopadhyay, and
  Chen]{cite727gal2014data}
Tamas~S Gal, Thomas~C Tucker, Aryya Gangopadhyay, and Zhiyuan Chen.
\newblock A data recipient centered de-identification method to retain
  statistical attributes.
\newblock \emph{Journal of biomedical informatics}, 50:\penalty0 32--45, 2014.

\bibitem[Ghinita et~al.(2008)Ghinita, Tao, and
  Kalnis]{cite728ghinita2008anonymization}
Gabriel Ghinita, Yufei Tao, and Panos Kalnis.
\newblock On the anonymization of sparse high-dimensional data.
\newblock In \emph{2008 IEEE 24th International Conference on Data
  Engineering}, pages 715--724. IEEE, 2008.

\bibitem[Gkoulalas-Divanis et~al.(2014)Gkoulalas-Divanis, Loukides, and
  Sun]{cite729gkoulalas2014publishing}
Aris Gkoulalas-Divanis, Grigorios Loukides, and Jimeng Sun.
\newblock Publishing data from electronic health records while preserving
  privacy: A survey of algorithms.
\newblock \emph{Journal of biomedical informatics}, 50:\penalty0 4--19, 2014.

\bibitem[Guntuku et~al.(2020)Guntuku, Sherman, Stokes, Agarwal, Seltzer,
  Merchant, and Ungar]{guntuku2020tracking}
Sharath~Chandra Guntuku, Garrick Sherman, Daniel~C Stokes, Anish~K Agarwal,
  Emily Seltzer, Raina~M Merchant, and Lyle~H Ungar.
\newblock Tracking mental health and symptom mentions on twitter during
  covid-19.
\newblock \emph{Journal of general internal medicine}, 35\penalty0
  (9):\penalty0 2798--2800, 2020.

\bibitem[Gyrard et~al.(2018)Gyrard, Gaur, Shekarpour, Thirunarayan, and
  Sheth]{gyrard2018personalized}
Amelia Gyrard, Manas Gaur, Saeedeh Shekarpour, Krishnaprasad Thirunarayan, and
  Amit Sheth.
\newblock Personalized health knowledge graph.
\newblock In \emph{ISWC 2018 Contextualized Knowledge Graph Workshop}, 2018.

\bibitem[Hamza et~al.(2013)Hamza, Hefny, et~al.]{hamza2013attacks}
Nermin Hamza, Hesham~A Hefny, et~al.
\newblock Attacks on anonymization-based privacy-preserving: a survey for data
  mining and data publishing.
\newblock \emph{Scientific Research Publishing}, 2013.

\bibitem[Hardt et~al.(2010)Hardt, Ligett, and McSherry]{hardt2010simple}
Moritz Hardt, Katrina Ligett, and Frank McSherry.
\newblock A simple and practical algorithm for differentially private data
  release.
\newblock \emph{arXiv preprint arXiv:1012.4763}, 2010.

\bibitem[Hay et~al.(2008)Hay, Miklau, Jensen, Towsley, and
  Weis]{hay2008resisting}
Michael Hay, Gerome Miklau, David Jensen, Don Towsley, and Philipp Weis.
\newblock Resisting structural re-identification in anonymized social networks.
\newblock \emph{Proceedings of the VLDB Endowment}, 1\penalty0 (1):\penalty0
  102--114, 2008.

\bibitem[J{\"a}ndel(2014)]{cite730jandel2014decision}
Magnus J{\"a}ndel.
\newblock Decision support for releasing anonymised data.
\newblock \emph{Computers \& security}, 46:\penalty0 48--61, 2014.

\bibitem[Ji et~al.(2016)Ji, Mittal, and Beyah]{ji2016graph}
Shouling Ji, Prateek Mittal, and Raheem Beyah.
\newblock Graph data anonymization, de-anonymization attacks, and
  de-anonymizability quantification: A survey.
\newblock \emph{IEEE Communications Surveys \& Tutorials}, 19\penalty0
  (2):\penalty0 1305--1326, 2016.

\bibitem[Jian-min et~al.(2008)Jian-min, Hui-qun, Juan, and
  Ting-ting]{jian2008complete}
Han Jian-min, Yu~Hui-qun, Yu~Juan, and Cen Ting-ting.
\newblock A complete (alpha, k)-anonymity model for sensitive values
  individuation preservation.
\newblock In \emph{2008 International Symposium on Electronic Commerce and
  Security}, pages 318--323. IEEE, 2008.

\bibitem[Johnson et~al.(2016)Johnson, Pollard, Shen, Li-Wei, Feng, Ghassemi,
  Moody, Szolovits, Celi, and Mark]{johnson2016mimic}
Alistair~EW Johnson, Tom~J Pollard, Lu~Shen, H~Lehman Li-Wei, Mengling Feng,
  Mohammad Ghassemi, Benjamin Moody, Peter Szolovits, Leo~Anthony Celi, and
  Roger~G Mark.
\newblock Mimic-iii, a freely accessible critical care database.
\newblock \emph{Scientific data}, 3\penalty0 (1):\penalty0 1--9, 2016.

\bibitem[Khan et~al.(2020)Khan, Tao, Anjum, Kanwal, Khan, Maple,
  et~al.]{khan2020theta}
Razaullah Khan, Xiaofeng Tao, Adeel Anjum, Tehsin Kanwal, Abid Khan, Carsten
  Maple, et~al.
\newblock $\theta$-sensitive k-anonymity: An anonymization model for iot based
  electronic health records.
\newblock \emph{Electronics}, 9\penalty0 (5):\penalty0 716, 2020.

\bibitem[Kieseberg et~al.(2016)Kieseberg, Malle, Fr{\"u}hwirt, Weippl, and
  Holzinger]{cite27kieseberg2016tamper}
Peter Kieseberg, Bernd Malle, Peter Fr{\"u}hwirt, Edgar Weippl, and Andreas
  Holzinger.
\newblock A tamper-proof audit and control system for the doctor in the loop.
\newblock \emph{Brain Informatics}, 3\penalty0 (4):\penalty0 269--279, 2016.

\bibitem[Kifer and Gehrke(2006)]{cite2kifer2006injecting}
Daniel Kifer and Johannes Gehrke.
\newblock Injecting utility into anonymized datasets.
\newblock In \emph{Proceedings of the 2006 ACM SIGMOD international conference
  on Management of data}, pages 217--228, 2006.

\bibitem[Kim et~al.(2018)Kim, Jang, and Yoo]{cite7kim2018privacy}
Jong~Wook Kim, Beakcheol Jang, and Hoon Yoo.
\newblock Privacy-preserving aggregation of personal health data streams.
\newblock \emph{PloS one}, 13\penalty0 (11):\penalty0 e0207639, 2018.

\bibitem[Kim et~al.(2014)Kim, Sung, and Chung]{cite79kim2014framework}
Soohyung Kim, Min~Kyoung Sung, and Yon~Dohn Chung.
\newblock A framework to preserve the privacy of electronic health data
  streams.
\newblock \emph{Journal of biomedical informatics}, 50:\penalty0 95--106, 2014.

\bibitem[Kisilevich et~al.(2009)Kisilevich, Rokach, Elovici, and
  Shapira]{cite444kisilevich2009efficient}
Slava Kisilevich, Lior Rokach, Yuval Elovici, and Bracha Shapira.
\newblock Efficient multidimensional suppression for k-anonymity.
\newblock \emph{IEEE Transactions on Knowledge and Data Engineering},
  22\penalty0 (3):\penalty0 334--347, 2009.

\bibitem[Lee and Chung(2020)]{lee2020differentially}
Hyukki Lee and Yon~Dohn Chung.
\newblock Differentially private release of medical microdata: an efficient and
  practical approach for preserving informative attribute values.
\newblock \emph{BMC Medical Informatics and Decision Making}, 20\penalty0
  (1):\penalty0 1--15, 2020.

\bibitem[Lee et~al.(2017)Lee, Kim, Kim, and Chung]{cite2lee2017utility}
Hyukki Lee, Soohyung Kim, Jong~Wook Kim, and Yon~Dohn Chung.
\newblock Utility-preserving anonymization for health data publishing.
\newblock \emph{BMC medical informatics and decision making}, 17\penalty0
  (1):\penalty0 1--12, 2017.

\bibitem[LeFevre et~al.(2006{\natexlab{a}})LeFevre, DeWitt, and
  Ramakrishnan]{cite432lefevre2006mondrian}
Kristen LeFevre, David~J DeWitt, and Raghu Ramakrishnan.
\newblock Mondrian multidimensional k-anonymity.
\newblock In \emph{22nd International conference on data engineering
  (ICDE'06)}, pages 25--25. IEEE, 2006{\natexlab{a}}.

\bibitem[LeFevre et~al.(2006{\natexlab{b}})LeFevre, DeWitt, and
  Ramakrishnan]{cite441lefevre2006workload}
Kristen LeFevre, David~J DeWitt, and Raghu Ramakrishnan.
\newblock Workload-aware anonymization.
\newblock In \emph{Proceedings of the 12th ACM SIGKDD international conference
  on Knowledge discovery and data mining}, pages 277--286, 2006{\natexlab{b}}.

\bibitem[Li et~al.(2015)Li, Xiong, Jiang, and Liu]{li2015differentially}
Haoran Li, Li~Xiong, Xiaoqian Jiang, and Jinfei Liu.
\newblock Differentially private histogram publication for dynamic datasets: an
  adaptive sampling approach.
\newblock In \emph{Proceedings of the 24th ACM International on Conference on
  Information and Knowledge Management}, pages 1001--1010, 2015.

\bibitem[Li et~al.(2011)Li, Liu, Baig, and Wong]{cite443li2011information}
Jiuyong Li, Jixue Liu, Muzammil Baig, and Raymond Chi-Wing Wong.
\newblock Information based data anonymization for classification utility.
\newblock \emph{Data \& Knowledge Engineering}, 70\penalty0 (12):\penalty0
  1030--1045, 2011.

\bibitem[Li et~al.(2007)Li, Li, and Venkatasubramanian]{cite3li2007t}
Ninghui Li, Tiancheng Li, and Suresh Venkatasubramanian.
\newblock t-closeness: Privacy beyond k-anonymity and l-diversity.
\newblock In \emph{2007 IEEE 23rd International Conference on Data
  Engineering}, pages 106--115. IEEE, 2007.

\bibitem[Li et~al.(2009)Li, Li, and Venkatasubramanian]{li2009closeness}
Ninghui Li, Tiancheng Li, and Suresh Venkatasubramanian.
\newblock Closeness: A new privacy measure for data publishing.
\newblock \emph{IEEE Transactions on Knowledge and Data Engineering},
  22\penalty0 (7):\penalty0 943--956, 2009.

\bibitem[Li and Li(2008)]{cite435li2008injector}
Tiancheng Li and Ninghui Li.
\newblock Injector: Mining background knowledge for data anonymization.
\newblock In \emph{2008 IEEE 24th International Conference on Data
  Engineering}, pages 446--455. IEEE, 2008.

\bibitem[Li et~al.(2010)Li, Li, Zhang, and Molloy]{cite2li2010slicing}
Tiancheng Li, Ninghui Li, Jian Zhang, and Ian Molloy.
\newblock Slicing: A new approach for privacy preserving data publishing.
\newblock \emph{IEEE transactions on knowledge and data engineering},
  24\penalty0 (3):\penalty0 561--574, 2010.

\bibitem[Liu et~al.(2016)Liu, Bier, Wilson, Guerra-Gomez, Honda, Sricharan,
  Gilpin, and Davies]{liu2016graph}
Juan Liu, Eric Bier, Aaron Wilson, John~Alexis Guerra-Gomez, Tomonori Honda,
  Kumar Sricharan, Leilani Gilpin, and Daniel Davies.
\newblock Graph analysis for detecting fraud, waste, and abuse in healthcare
  data.
\newblock \emph{AI Magazine}, 37\penalty0 (2):\penalty0 33--46, 2016.

\bibitem[Liu and Terzi(2008)]{cite58liu2008towards}
Kun Liu and Evimaria Terzi.
\newblock Towards identity anonymization on graphs.
\newblock In \emph{Proceedings of the 2008 ACM SIGMOD international conference
  on Management of data}, pages 93--106, 2008.

\bibitem[Liu et~al.(2018)Liu, Dou, Chen, Olatunji, Qin, and Heng]{liu2018mtmr}
Lihao Liu, Qi~Dou, Hao Chen, Iyiola~E Olatunji, Jing Qin, and Pheng-Ann Heng.
\newblock Mtmr-net: Multi-task deep learning with margin ranking loss for lung
  nodule analysis.
\newblock In \emph{Deep Learning in Medical Image Analysis and Multimodal
  Learning for Clinical Decision Support}, pages 74--82. Springer, 2018.

\bibitem[Machanavajjhala et~al.(2007)Machanavajjhala, Kifer, Gehrke, and
  Venkitasubramaniam]{cite3machanavajjhala2007diversity}
Ashwin Machanavajjhala, Daniel Kifer, Johannes Gehrke, and Muthuramakrishnan
  Venkitasubramaniam.
\newblock l-diversity: Privacy beyond k-anonymity.
\newblock \emph{ACM Transactions on Knowledge Discovery from Data (TKDD)},
  1\penalty0 (1):\penalty0 3--es, 2007.

\bibitem[Majeed(2019)]{cite3majeed2019attribute}
Abdul Majeed.
\newblock Attribute-centric anonymization scheme for improving user privacy and
  utility of publishing e-health data.
\newblock \emph{Journal of King Saud University-Computer and Information
  Sciences}, 31\penalty0 (4):\penalty0 426--435, 2019.

\bibitem[Majeed and Lee(2020)]{majeed2020anonymization}
Abdul Majeed and Sungchang Lee.
\newblock Anonymization techniques for privacy preserving data publishing: A
  comprehensive survey.
\newblock \emph{IEEE Access}, 2020.

\bibitem[Majeed et~al.(2017)Majeed, Ullah, and
  Lee]{cite4majeed2017vulnerability}
Abdul Majeed, Farman Ullah, and Sungchang Lee.
\newblock Vulnerability-and diversity-aware anonymization of personally
  identifiable information for improving user privacy and utility of publishing
  data.
\newblock \emph{Sensors}, 17\penalty0 (5):\penalty0 1059, 2017.

\bibitem[McSherry(2009)]{cite415mcsherry2009privacy}
Frank~D McSherry.
\newblock Privacy integrated queries: an extensible platform for
  privacy-preserving data analysis.
\newblock In \emph{Proceedings of the 2009 ACM SIGMOD International Conference
  on Management of data}, pages 19--30, 2009.

\bibitem[Mohammadian et~al.(2014)Mohammadian, Noferesti, and
  Jalili]{cite7mohammadian2014fast}
Esmaeil Mohammadian, Morteza Noferesti, and Rasool Jalili.
\newblock Fast: fast anonymization of big data streams.
\newblock In \emph{Proceedings of the 2014 International Conference on Big Data
  Science and Computing}, pages 1--8, 2014.

\bibitem[Mohammed et~al.(2010)Mohammed, Fung, Hung, and
  Lee]{cite732mohammed2010centralized}
Noman Mohammed, Benjamin~CM Fung, Patrick~CK Hung, and Cheuk-Kwong Lee.
\newblock Centralized and distributed anonymization for high-dimensional
  healthcare data.
\newblock \emph{ACM Transactions on Knowledge Discovery from Data (TKDD)},
  4\penalty0 (4):\penalty0 1--33, 2010.

\bibitem[Mohan et~al.(2012)Mohan, Thakurta, Shi, Song, and
  Culler]{cite417mohan2012gupt}
Prashanth Mohan, Abhradeep Thakurta, Elaine Shi, Dawn Song, and David Culler.
\newblock Gupt: privacy preserving data analysis made easy.
\newblock In \emph{Proceedings of the 2012 ACM SIGMOD International Conference
  on Management of Data}, pages 349--360, 2012.

\bibitem[Mortazavi and Erfani(2020)]{mortazavi2020gram}
R~Mortazavi and SH~Erfani.
\newblock Gram: An efficient (k, l) graph anonymization method.
\newblock \emph{Expert Systems with Applications}, 153:\penalty0 113454, 2020.

\bibitem[Mortazavi and Jalili(2014)]{cite733mortazavi2014fast}
Reza Mortazavi and Saeed Jalili.
\newblock Fast data-oriented microaggregation algorithm for large numerical
  datasets.
\newblock \emph{Knowledge-Based Systems}, 67:\penalty0 195--205, 2014.

\bibitem[Narayanan and Shmatikov(2008)]{cite734narayanan2008robust}
Arvind Narayanan and Vitaly Shmatikov.
\newblock Robust de-anonymization of large sparse datasets.
\newblock In \emph{2008 IEEE Symposium on Security and Privacy (sp 2008)},
  pages 111--125. IEEE, 2008.

\bibitem[Nergiz and G{\"o}k(2014)]{cite29nergiz2014hybrid}
Mehmet~Ercan Nergiz and Muhammed~Zahit G{\"o}k.
\newblock Hybrid k-anonymity.
\newblock \emph{Computers \& security}, 44:\penalty0 51--63, 2014.

\bibitem[Nergiz et~al.(2008)Nergiz, Clifton, and
  Nergiz]{cite735nergiz2008multirelational}
Mehmet~Ercan Nergiz, Christopher Clifton, and Ahmet~Erhan Nergiz.
\newblock Multirelational k-anonymity.
\newblock \emph{IEEE Transactions on Knowledge and Data Engineering},
  21\penalty0 (8):\penalty0 1104--1117, 2008.

\bibitem[Oganian and Domingo-Ferrer(2001)]{cite5oganian2001complexity}
Anna Oganian and Josep Domingo-Ferrer.
\newblock On the complexity of optimal microaggregation for statistical
  disclosure control.
\newblock \emph{Statistical Journal of the United Nations Economic Commission
  for Europe}, 18\penalty0 (4):\penalty0 345--353, 2001.

\bibitem[Oishi et~al.(2020)Oishi, Sei, Tahara, and Ohsuga]{oishi2020semantic}
Keiichiro Oishi, Yuichi Sei, Yasuyuki Tahara, and Akihiko Ohsuga.
\newblock Semantic diversity: Privacy considering distance between values of
  sensitive attribute.
\newblock \emph{Computers \& Security}, 94:\penalty0 101823, 2020.

\bibitem[Olatunji et~al.(2021)Olatunji, Nejdl, and
  Khosla]{olatunji2021membership}
Iyiola~E Olatunji, Wolfgang Nejdl, and Megha Khosla.
\newblock Membership inference attack on graph neural networks.
\newblock \emph{arXiv preprint arXiv:2101.06570}, 2021.

\bibitem[Otgonbayar et~al.(2018)Otgonbayar, Pervez, Dahal, and
  Eager]{cite7otgonbayar2018k}
Ankhbayar Otgonbayar, Zeeshan Pervez, Keshav Dahal, and Steve Eager.
\newblock K-varp: K-anonymity for varied data streams via partitioning.
\newblock \emph{Information Sciences}, 467:\penalty0 238--255, 2018.

\bibitem[Otgonbayar et~al.(2019)Otgonbayar, Pervez, and
  Dahal]{cite7otgonbayar2019x}
Ankhbayar Otgonbayar, Zeeshan Pervez, and Keshav Dahal.
\newblock $ x-band $: Expiration band for anonymizing varied data streams.
\newblock \emph{IEEE Internet of Things Journal}, 7\penalty0 (2):\penalty0
  1438--1450, 2019.

\bibitem[Pakbin et~al.(2018)Pakbin, Rafi, Hurley, Schulz, Krumholz, and
  Mortazavi]{pakbin2018prediction}
Arash Pakbin, Parvez Rafi, Nate Hurley, Wade Schulz, M~Harlan Krumholz, and
  J~Bobak Mortazavi.
\newblock Prediction of icu readmissions using data at patient discharge.
\newblock In \emph{2018 40th Annual International Conference of the IEEE
  Engineering in Medicine and Biology Society (EMBC)}, pages 4932--4935. IEEE,
  2018.

\bibitem[Parliament and of~the European~Union(2016)]{officialgrpr2016}
European Parliament and Council of~the European~Union.
\newblock General data protection regulation.
\newblock \emph{Official Journal of the European Union}, 2016.
\newblock URL
  \url{https://eur-lex.europa.eu/legal-content/EN/ALL/?uri=CELEX:32016R0679}.

\bibitem[Pei et~al.(2007)Pei, Xu, Wang, Wang, and
  Wang]{cite737pei2007maintaining}
Jian Pei, Jian Xu, Zhibin Wang, Wei Wang, and Ke~Wang.
\newblock Maintaining k-anonymity against incremental updates.
\newblock In \emph{19th International Conference on Scientific and Statistical
  Database Management (SSDBM 2007)}, pages 5--5. IEEE, 2007.

\bibitem[Poulis et~al.(2017)Poulis, Loukides, Skiadopoulos, and
  Gkoulalas-Divanis]{POULIS201776}
Giorgos Poulis, Grigorios Loukides, Spiros Skiadopoulos, and Aris
  Gkoulalas-Divanis.
\newblock Anonymizing datasets with demographics and diagnosis codes in the
  presence of utility constraints.
\newblock \emph{Journal of Biomedical Informatics}, 65:\penalty0 76--96, 2017.
\newblock ISSN 1532-0464.
\newblock \doi{https://doi.org/10.1016/j.jbi.2016.11.001}.

\bibitem[Prasser et~al.(2020)Prasser, Eicher, Spengler, Bild, and
  Kuhn]{prasser2020flexible}
Fabian Prasser, Johanna Eicher, Helmut Spengler, Raffael Bild, and Klaus~A
  Kuhn.
\newblock Flexible data anonymization using arx—current status and challenges
  ahead.
\newblock \emph{Software: Practice and Experience}, 50\penalty0 (7):\penalty0
  1277--1304, 2020.

\bibitem[Qu et~al.(2017)Qu, Xu, and Yu]{qu2017privacy}
Youyang Qu, Jiyang Xu, and Shui Yu.
\newblock Privacy preserving in big data sets through multiple shuffle.
\newblock In \emph{Proceedings of the Australasian Computer Science Week
  Multiconference}, pages 1--8, 2017.

\bibitem[Roy et~al.(2010)Roy, Setty, Kilzer, Shmatikov, and
  Witchel]{cite416roy2010airavat}
Indrajit Roy, Srinath~TV Setty, Ann Kilzer, Vitaly Shmatikov, and Emmett
  Witchel.
\newblock Airavat: Security and privacy for mapreduce.
\newblock In \emph{NSDI}, volume~10, pages 297--312, 2010.

\bibitem[Ryffel et~al.(2018)Ryffel, Trask, Dahl, Wagner, Mancuso, Rueckert, and
  Passerat-Palmbach]{ryffel2018genericPysyft}
Theo Ryffel, Andrew Trask, Morten Dahl, Bobby Wagner, Jason Mancuso, Daniel
  Rueckert, and Jonathan Passerat-Palmbach.
\newblock A generic framework for privacy preserving deep learning.
\newblock \emph{arXiv preprint arXiv:1811.04017}, 2018.

\bibitem[Sarathy and Muralidhar(2011)]{cite414sarathy2011evaluating}
Rathindra Sarathy and Krishnamurty Muralidhar.
\newblock Evaluating laplace noise addition to satisfy differential privacy for
  numeric data.
\newblock \emph{Trans. Data Priv.}, 4\penalty0 (1):\penalty0 1--17, 2011.

\bibitem[Sei et~al.(2017)Sei, Okumura, Takenouchi, and
  Ohsuga]{sei2017anonymization}
Yuichi Sei, Hiroshi Okumura, Takao Takenouchi, and Akihiko Ohsuga.
\newblock Anonymization of sensitive quasi-identifiers for l-diversity and
  t-closeness.
\newblock \emph{IEEE transactions on dependable and secure computing},
  16\penalty0 (4):\penalty0 580--593, 2017.

\bibitem[Shi et~al.(2017)Shi, Li, Yang, Qi, Pan, and Zhou]{shi2017semantic}
Longxiang Shi, Shijian Li, Xiaoran Yang, Jiaheng Qi, Gang Pan, and Binbin Zhou.
\newblock Semantic health knowledge graph: semantic integration of
  heterogeneous medical knowledge and services.
\newblock \emph{BioMed research international}, 2017, 2017.

\bibitem[Soria-Comas et~al.(2014)Soria-Comas, Domingo-Ferrer, S{\'a}nchez, and
  Mart{\'\i}nez]{soria2014enhancing}
Jordi Soria-Comas, Josep Domingo-Ferrer, David S{\'a}nchez, and Sergio
  Mart{\'\i}nez.
\newblock Enhancing data utility in differential privacy via
  microaggregation-based k-anonymity.
\newblock \emph{The VLDB Journal}, 23\penalty0 (5):\penalty0 771--794, 2014.

\bibitem[Soria-Comas et~al.(2015)Soria-Comas, Domingo-Ferrer, Sanchez, and
  Martinez]{soria2015t}
Jordi Soria-Comas, Josep Domingo-Ferrer, David Sanchez, and Sergio Martinez.
\newblock t-closeness through microaggregation: Strict privacy with enhanced
  utility preservation.
\newblock \emph{IEEE Transactions on Knowledge and Data Engineering},
  27\penalty0 (11):\penalty0 3098--3110, 2015.

\bibitem[Soria-Comas et~al.(2017)Soria-Comas, Domingo-Ferrer, S{\'a}nchez, and
  Meg{\'\i}as]{cite413soria2017individual}
Jordi Soria-Comas, Josep Domingo-Ferrer, David S{\'a}nchez, and David
  Meg{\'\i}as.
\newblock Individual differential privacy: A utility-preserving formulation of
  differential privacy guarantees.
\newblock \emph{IEEE Transactions on Information Forensics and Security},
  12\penalty0 (6):\penalty0 1418--1429, 2017.

\bibitem[Sun et~al.(2011)Sun, Sun, and Wang]{sun2011extended}
Xiaoxun Sun, Lili Sun, and Hua Wang.
\newblock Extended k-anonymity models against sensitive attribute disclosure.
\newblock \emph{Computer Communications}, 34\penalty0 (4):\penalty0 526--535,
  2011.

\bibitem[Sweeney(2000)]{sweeney2000simple}
Latanya Sweeney.
\newblock Simple demographics often identify people uniquely.
\newblock \emph{Health (San Francisco)}, 671\penalty0 (2000):\penalty0 1--34,
  2000.

\bibitem[Sweeney(2002)]{cite28sweeney2002achieving}
Latanya Sweeney.
\newblock Achieving k-anonymity privacy protection using generalization and
  suppression.
\newblock \emph{International Journal of Uncertainty, Fuzziness and
  Knowledge-Based Systems}, 10\penalty0 (05):\penalty0 571--588, 2002.

\bibitem[Tang et~al.(2018)Tang, Xiao, Wang, and Zhou]{tang2018predictive}
Fengyi Tang, Cao Xiao, Fei Wang, and Jiayu Zhou.
\newblock Predictive modeling in urgent care: a comparative study of machine
  learning approaches.
\newblock \emph{Jamia Open}, 1\penalty0 (1):\penalty0 87--98, 2018.

\bibitem[Tao et~al.(2008)Tao, Xiao, Li, and Zhang]{cite438tao2008anti}
Yufei Tao, Xiaokui Xiao, Jiexing Li, and Donghui Zhang.
\newblock On anti-corruption privacy preserving publication.
\newblock In \emph{2008 IEEE 24th International Conference on Data
  Engineering}, pages 725--734. IEEE, 2008.

\bibitem[Thompson and Yao(2009)]{thompson2009union}
Brian Thompson and Danfeng Yao.
\newblock The union-split algorithm and cluster-based anonymization of social
  networks.
\newblock In \emph{Proceedings of the 4th International Symposium on
  Information, Computer, and Communications Security}, pages 218--227, 2009.

\bibitem[Tian and Zhang(2011)]{tian2011extending}
Hongwei Tian and Weining Zhang.
\newblock Extending l-diversity to generalize sensitive data.
\newblock \emph{Data \& Knowledge Engineering}, 70\penalty0 (1):\penalty0
  101--126, 2011.

\bibitem[Walonoski et~al.(2017)Walonoski, Kramer, Nichols, Quina, Moesel, Hall,
  Duffett, Dube, Gallagher, and McLachlan]{Synthea10.1093/jamia/ocx079}
Jason Walonoski, Mark Kramer, Joseph Nichols, Andre Quina, Chris Moesel, Dylan
  Hall, Carlton Duffett, Kudakwashe Dube, Thomas Gallagher, and Scott
  McLachlan.
\newblock {Synthea: An approach, method, and software mechanism for generating
  synthetic patients and the synthetic electronic health care record}.
\newblock \emph{Journal of the American Medical Informatics Association},
  25\penalty0 (3):\penalty0 230--238, 08 2017.
\newblock ISSN 1527-974X.
\newblock \doi{10.1093/jamia/ocx079}.
\newblock URL \url{https://doi.org/10.1093/jamia/ocx079}.

\bibitem[Wang et~al.(2018)Wang, Zhu, Chen, and Chang]{wang2018privacy}
Rong Wang, Yan Zhu, Tung-Shou Chen, and Chin-Chen Chang.
\newblock Privacy-preserving algorithms for multiple sensitive attributes
  satisfying t-closeness.
\newblock \emph{Journal of Computer Science and Technology}, 33\penalty0
  (6):\penalty0 1231--1242, 2018.

\bibitem[Wong et~al.(2006)Wong, Li, Fu, and Wang]{wong2006alpha}
Raymond Chi-Wing Wong, Jiuyong Li, Ada Wai-Chee Fu, and Ke~Wang.
\newblock ($\alpha$, k)-anonymity: an enhanced k-anonymity model for privacy
  preserving data publishing.
\newblock In \emph{Proceedings of the 12th ACM SIGKDD international conference
  on Knowledge discovery and data mining}, pages 754--759, 2006.

\bibitem[Wong et~al.(2007)Wong, Fu, Wang, and Pei]{cite437wong2007minimality}
Raymond Chi-Wing Wong, Ada Wai-Chee Fu, Ke~Wang, and Jian Pei.
\newblock Minimality attack in privacy preserving data publishing.
\newblock In \emph{Proceedings of the 33rd international conference on Very
  large data bases}, pages 543--554, 2007.

\bibitem[Wu et~al.(2019)Wu, Wei, Jiang, Wang, and Jiang]{wu2019micro}
Xiang Wu, Yuyang Wei, Tao Jiang, Yu~Wang, and Shuguang Jiang.
\newblock A micro-aggregation algorithm based on density partition method for
  anonymizing biomedical data.
\newblock \emph{Current Bioinformatics}, 14\penalty0 (7):\penalty0 667--675,
  2019.

\bibitem[Wu et~al.(2009)Wu, Sun, and Wang]{cite739wu2009privacy}
Yingjie Wu, Zhihui Sun, and Xiaodong Wang.
\newblock Privacy preserving k-anonymity for re-publication of incremental
  datasets.
\newblock In \emph{2009 WRI World Congress on Computer Science and Information
  Engineering}, volume~4, pages 53--60. IEEE, 2009.

\bibitem[Xiao and Li(2020)]{xiao2020privacy}
Yuelei Xiao and Haiqi Li.
\newblock Privacy preserving data publishing for multiple sensitive attributes
  based on security level.
\newblock \emph{Information}, 11\penalty0 (3):\penalty0 166, 2020.

\bibitem[Xu et~al.(2006)Xu, Wang, Pei, Wang, Shi, and Fu]{cite433xu2006utility}
Jian Xu, Wei Wang, Jian Pei, Xiaoyuan Wang, Baile Shi, and Ada Wai-Chee Fu.
\newblock Utility-based anonymization using local recoding.
\newblock In \emph{Proceedings of the 12th ACM SIGKDD international conference
  on Knowledge discovery and data mining}, pages 785--790, 2006.

\bibitem[Yu et~al.(2017)Yu, Li, Yu, Tian, Shun, Xu, Zhu, and
  Gao]{yu2017knowledge}
Tong Yu, Jinghua Li, Qi~Yu, Ye~Tian, Xiaofeng Shun, Lili Xu, Ling Zhu, and
  Hongjie Gao.
\newblock Knowledge graph for tcm health preservation: design, construction,
  and applications.
\newblock \emph{Artificial intelligence in medicine}, 77:\penalty0 48--52,
  2017.

\bibitem[Yuan et~al.(2010)Yuan, Chen, and Yu]{cite5yuan2010personalized}
Mingxuan Yuan, Lei Chen, and Philip~S Yu.
\newblock Personalized privacy protection in social networks.
\newblock \emph{Proceedings of the VLDB Endowment}, 4\penalty0 (2):\penalty0
  141--150, 2010.

\bibitem[Zaman et~al.(2016)Zaman, Obimbo, and Dara]{cite427zaman2016novel}
ANK Zaman, Charlie Obimbo, and Rozita~A Dara.
\newblock A novel differential privacy approach that enhances classification
  accuracy.
\newblock In \emph{Proceedings of the Ninth International C* Conference on
  Computer Science \& Software Engineering}, pages 79--84, 2016.

\bibitem[Zhang et~al.(2017{\natexlab{a}})Zhang, Cormode, Procopiuc, Srivastava,
  and Xiao]{zhang2017privbayes}
Jun Zhang, Graham Cormode, Cecilia~M Procopiuc, Divesh Srivastava, and Xiaokui
  Xiao.
\newblock Privbayes: Private data release via bayesian networks.
\newblock \emph{ACM Transactions on Database Systems (TODS)}, 42\penalty0
  (4):\penalty0 1--41, 2017{\natexlab{a}}.

\bibitem[Zhang et~al.(2017{\natexlab{b}})Zhang, Xuan, Si, and
  Wang]{zhang2017improved}
Lin Zhang, Jie Xuan, Ruoqian Si, and Ruchuan Wang.
\newblock An improved algorithm of individuation k-anonymity for multiple
  sensitive attributes.
\newblock \emph{Wireless Personal Communications}, 95\penalty0 (3):\penalty0
  2003--2020, 2017{\natexlab{b}}.

\bibitem[Zhang et~al.(2007)Zhang, Koudas, Srivastava, and
  Yu]{zhang2007aggregate}
Qing Zhang, Nick Koudas, Divesh Srivastava, and Ting Yu.
\newblock Aggregate query answering on anonymized tables.
\newblock In \emph{2007 IEEE 23rd international conference on data
  engineering}, pages 116--125. IEEE, 2007.

\bibitem[Zheleva and Getoor(2007)]{cite517zheleva2007preserving}
Elena Zheleva and Lise Getoor.
\newblock Preserving the privacy of sensitive relationships in graph data.
\newblock In \emph{International workshop on privacy, security, and trust in
  KDD}, pages 153--171. Springer, 2007.

\bibitem[Zhou and Pei(2008)]{cite518zhou2008preserving}
Bin Zhou and Jian Pei.
\newblock Preserving privacy in social networks against neighborhood attacks.
\newblock In \emph{2008 IEEE 24th International Conference on Data
  Engineering}, pages 506--515. IEEE, 2008.

\bibitem[Zhu et~al.(2015)Zhu, Tian, and L{\"u}]{zhu2015privacy}
Hong Zhu, Shengli Tian, and Kevin L{\"u}.
\newblock Privacy-preserving data publication with features of independent
  l-diversity.
\newblock \emph{The Computer Journal}, 58\penalty0 (4):\penalty0 549--571,
  2015.

\bibitem[Zou et~al.(2009)Zou, Chen, and {\"O}zsu]{cite5zou2009k}
Lei Zou, Lei Chen, and M~Tamer {\"O}zsu.
\newblock K-automorphism: A general framework for privacy preserving network
  publication.
\newblock \emph{Proceedings of the VLDB Endowment}, 2\penalty0 (1):\penalty0
  946--957, 2009.

\end{thebibliography}

\end{document}